\documentclass[twocolumn,showpacs,preprintnumbers,amsmath,amssymb]{revtex4}

\usepackage{epsfig}
\usepackage{subfigure}
\usepackage{bm}
\usepackage{graphicx}
\usepackage{dcolumn}
\usepackage{epsfig,amsmath}
\graphicspath{{pics/}}    

\usepackage[normalem]{ulem} 
\usepackage[dvips]{color} 
\renewcommand\sout{\bgroup \color{red} \ULdepth=-.5ex \ULset}

\newcommand{\be}{\begin{equation}}
\newcommand{\ee}{\end{equation}}
\newcommand{\bea}{\begin{eqnarray}}
\newcommand{\eea}{\end{eqnarray}}

\newcommand{\UNIT}[1]{\mbox{$\,{\rm #1}$}}

\newcommand{\fm}{\UNIT{fm}}

\def\NP{N_\mathrm{Part}}

\def\bi{\begin{itemize}}
\def\ei{\end{itemize}}
\def\D{$D$}
\def\d{{\rm d}}
\def\e{{\rm e}}

\def\NPt{$\mathrm{N}_\mathrm{Part}$}

\def\fm3{$\mathrm{fm}^3$}
\def\f3{\mathrm{fm}^3}

\def\be{\begin{equation}}
\def\ee{\end{equation}}

\def\jpsi{J/$\psi$}

\def\sqrtsnn{$\sqrt{s}_\mathrm{NN}$}

\def\sigmabb{\sigma_{\mathrm{\bar{b}b}}}
\def\sigmacc{\sigma_{\mathrm{\bar{c}c}}}
\def\B0bar{$\bar{B^0}$}

\begin{document}

\title{Dilepton production in proton-proton and Pb+Pb collisions at $\sqrt{s_\mathrm{NN}}$  = 2.76 TeV}

\author{O.~Linnyk}

\email{Olena.Linnyk@theo.physik.uni-giessen.de}

\affiliation{%
 Institut f\"ur Theoretische Physik, %
  Universit\"at Giessen, %
  35392 Giessen, %
  Germany %
}

\author{J.~Manninen}%
\affiliation{%
 Institut f\"ur Theoretische Physik, %
 Johann Wolfgang Goethe University, %
 60438 Frankfurt am Main, %
 Germany; %
Frankfurt Institute for Advanced Studies, %
 60438 Frankfurt am Main, %
 Germany; %
}

\author{E.~L.~Bratkovskaya}%
\affiliation{%
 Institut f\"ur Theoretische Physik, %
 Johann Wolfgang Goethe University, %
 60438 Frankfurt am Main, %
 Germany; %
Frankfurt Institute for Advanced Studies, %
 60438 Frankfurt am Main, %
 Germany; %
}

\author{P.~B.~Gossiaux}%
\affiliation{%
SUBATECH, UMR 6457, Laboratoire de Physique Subatomique et des
Technologies Associ\'ees, %
University of Nantes - IN2P3/CNRS - Ecole des Mines de Nantes, %
44072 Nantes Cedex 03, %
France; }

\author{J.~Aichelin}%
\affiliation{%
SUBATECH, UMR 6457, Laboratoire de Physique Subatomique et des
Technologies Associ\'ees, %
University of Nantes - IN2P3/CNRS - Ecole des Mines de Nantes, %
44072 Nantes Cedex 03, %
France; }

\author{W.~Cassing}
\affiliation{%
 Institut f\"ur Theoretische Physik, %
  Universit\"at Giessen, %
  35392 Giessen, %
  Germany %
}

\author{T.~Song}%
\affiliation{%
Cyclotron Institute and Department of Physics and Astronomy, %
Texas A\&M University, %
College Station, TX 77843-3366, USA}

\author{C.~M.~Ko}%
\affiliation{%
Cyclotron Institute and Department of Physics and Astronomy, %
Texas A\&M University, %
College Station, TX 77843-3366, USA}

\date{\today}

\begin{abstract}
We study $e^+e^-$ pair production in proton-proton and central Pb+Pb
collisions at $\sqrt{s_\mathrm{NN}}= 2.76$~TeV within two models: an
extended statistical hadronization model (SHM) and the
Parton-Hadron-String Dynamics (PHSD) transport approach. We find
that the PHSD calculations roughly agree with the dilepton spectrum
from hadronic sources with the `cocktail' estimates from the
statistical hadronization model matched to available data at LHC
energies. The dynamical simulations within the PHSD show a moderate
increase of the low mass dilepton yield essentially due to the
in-medium modification of the $\rho$-meson. Furthermore, pronounced
traces of the partonic degrees of freedom are found in the PHSD
results in the intermediate mass regime. The dilepton production
from the strongly interacting quark-gluon plasma (sQGP) exceeds that
from the semi-leptonic decays of open charm and bottom mesons.
Additionally, we observe that a transverse momentum cut of
1~GeV/{$c$} further suppresses the relative contribution of the
heavy meson decays to the dilepton yield, such that the sQGP
radiation strongly dominates the spectrum for masses from 1 to
3~GeV, allowing a closer look at the electromagnetic emissivity of
the partonic plasma in the early phase of Pb+Pb collisions.
\end{abstract}


\maketitle

\section{Introduction}

Dileptons, i.e. correlated  electron and positron or $\mu^+ \mu^-$
pairs, are one of the key observables in {experiments for}
ultra-relativistic nuclear collisions
 since {they} are emitted during the whole evolution
 {of a collision} and {interact only electromagnetically
 and thus very weakly with the strongly interacting partonic or
 hadronic medium created in the collisions.}
 {Also, dileptons of different invariant masses are dominantly produced from}
 different stages of a relativistic nuclear collision{,
 and this provides the possibility to probe the properties of the
 produced hot and dense matter at various conditions} by measuring
 {the} differential dilepton spectra.

{T}he invariant mass spectrum {of dileptons} can be roughly divided
into 3 different regions which are dominated by different physics.
In the low mass region ($M_{\e^+\e^-}<$ 1 GeV){,} the radiation is
dominated by the decays of light mesons (consisting of $u$, $d$ and
$s$ (anti)quarks){, while in}  the intermediate mass region (1 GeV
$<M_{\e^+\e^-}<$ 3 GeV){,} the dominant hadronic contribution to the
invariant mass spectrum stems from the decays of open charm
mesons{.} {A}bove the $J/\psi$ peak, { the dilepton spectrum is}
first {dominated by} open beauty decays and {then by the} initial
state Drell-Yan radiation. {Besides these} sources {of dileptons},
the radiation from the strongly interacting Quark-Gluon-Plasma
(sQGP)~\cite{Shuryak:1978ij} as well as some other more exotic
sources like simultaneous interactions of four
pions~\cite{Song:1994zs,Li:1998ma,vanHees:2006ng,vanHees:2007th}
{can also give significant contributions,  particularly in the
intermediate mass region.} These partonic and hadronic channels have
been studied in detail in Refs. \cite{Linnyk:2011vx,Linnyk:2011hz}
at top Super-Proton-Synchrotron (SPS) and
Relativistic-Heavy-Ion-Collider (RHIC) energies{,} and it has been
found that the partonic channels clearly dominate over multi-pion
sources in the intermediate dilepton mass regime.

{In the present} study{,} {we}  include all known (leading) dilepton
sources to study and compare the magnitude of the {radiation from
the} QGP with {that from} other (conventional) sources {in heavy ion
collisions} at {the} Large-Hadron-Collider (LHC) energies. {Since at
present} there is no single model that could address reliably all of
the above-mentioned sources {for dilpeton production,} we {thus}
employ in this analysis different approaches to evaluate the
invariant mass spectrum from the different sources. {Specifically,
we} concentrate on two models (and their extensions): The
Parton-Hadron-String-Dynamics (PHSD) approach~\cite{CasBrat,BrCa11}
and the extended Statistical Hadronization Model
(SHM)~\cite{Manninen:2010yf}. The PHSD model is a relativistic
transport approach {developed}, tested and well suited for
{studying} dynamical partonic and hadronic systems {in heavy ion
collisions} from low to ultra-relativistic energies of
$\sqrt{s_\mathrm{NN}}$=200 GeV{,} while the SHM is a statistical
hadronic model {suitable} for {describing} in detail the relative
yields of final state hadrons.

In the case of proton-proton ($pp$) collisions, {they are dominated
by} the creation and decays of various hadrons{,} and the SHM should
be a suitable model {for estimating} the hadronic freeze-out
'cocktail' contribution {to dileptons}. Indeed, {this was confirmed
by} our simulations in Ref.~\cite{Manninen:2010yf} for p+p
collisions at $\sqrt{s_\mathrm{NN}}$= 200 GeV. On the contrary, in
heavy-ion collisions a rather long-living strongly interacting
dynamical fireball is formed and dileptons are emitted from the
created charges over an extended period of time. For {such} systems
the SHM might not provide a good description because the properties,
i.e. the masses and widths, of hadrons might change in the hot and
dense nuclear environment as a function of time, which can not be
taken into account in the SHM. Additionally, the dilepton signal
{from the QGP during} the early reaction phase is expected {to
outshine the background radiation due to} the cocktail of hadron
decays. For this purpose, we will employ the PHSD transport approach
which can account for the medium-dependent properties of partons and
hadrons and {their} multiple interactions as well as for the
dynamical evolution of the system in general, including the
transition to the partonic phase and the radiation from the sQGP.

We note that a direct comparison between results from the PHSD and
the SHM has shown that the low mass ($M<1$~GeV) spectra of
$\e^+\e^-$ -pairs in $Au$+$Au$ collisions at \sqrtsnn = 200 GeV
deviate from each other by only up to
20\%~\cite{Manninen:2010yf,Linnyk:2011hz}, i.e. either the
time-evolution of the system plays only a minor role in the dilepton
emission or the subsequent hadronic evolution to a large extent
washes out the details of the dynamics in the dilepton invariant
mass spectrum. We will study this issue again in detail in the
{first part of present study} and compare explicitly our {results
from both the} static and dynamical {model} in order to constrain
the dilepton yield from known hadronic sources. {We} will {then}
proceed to our main goal, i.e. the identification of the QGP signal
{from} the spectrum of dileptons produced in $Pb+Pb$ collisions in
the LHC energy regime.

In the following Section (Sect.~\ref{section_phsd}){,} we will
briefly recall the main concepts of the PHSD transport approach and
list the partonic and hadronic sources of dileptons in PHSD. Next,
we describe the SHM and the way hadron yields and spectra are
evaluated
 in Sect.~\ref{section_shm}. Our treatment of $D$ and $B$ meson
production and energy loss is described in
Sect.~\ref{sectioncharm}{,} whereas the modeling of charmonia
production in elementary and in nucleus-nucleus collisions is
outlined in Sect.\ref{section_cc}.  {R}esults {from our study} are
presented in Sect.\ref{sectionresults}{. We first include a
discussion of dilepton production in} the baseline p+p collisions{.}
We {then} continue with results from the SHM as well as from the
dynamical calculations of Pb+Pb collisions within the PHSD.
Further{more,} the effect of in-medium modifications of vector
mesons on the low-mass dilepton spectrum at
$\sqrt{s_\mathrm{NN}}$=2.76 TeV is investigated. {F}inally, the
influence of a transverse momentum cut on the {dilepton spectrum as
a possible way to enhance the} QGP signal is explored.  We conclude
our study in Sect.~\ref{sectionsummary} with a summary of our
findings and a discussion of open problems.

\section{The Parton-Hadron-String Dynamics (PHSD) transport approach}
\label{section_phsd}

The {PHSD} model~\cite{CasBrat,BrCa11} is an off-shell transport
model that consistently describes the full evolution of a
relativistic heavy-ion collision from the initial hard scatterings
and string formation through the dynamical deconfinement phase
transition to the quark-gluon plasma as well as hadronization and to
the subsequent interactions in the hadronic phase. In the hadronic
sector, the PHSD is equivalent to the Hadron-String-Dynamics (HSD)
transport approach~\cite{Cass99,Brat97,Ehehalt} that has been used
for the description of $pA$ and $AA$ collisions from SIS to RHIC
energies and has led to a fair reproduction of measured hadron
abundances, rapidity distributions and transverse momentum spectra.
In particular, {as in the} HSD, the PHSD incorporates off-shell
dynamics for vector mesons~\cite{Cass_off1} and a set of
vector-meson spectral functions~\cite{Brat08} that covers possible
scenarios for their in-medium modifications. In PHSD the transition
from the partonic to hadronic degrees of freedom is described by
covariant transition rates for the fusion of quark-antiquark pairs
to mesonic resonances or three quarks (antiquarks) to baryonic
states, i.e., by {the} dynamical hadronization \cite{Cass08}. Note
that due to the off-shell nature of both partons and hadrons, the
hadronization process obeys all conservation laws (i.e., the
4-momentum conservation and the flavor current conservation) in each
event, the detailed balance relations, and the increase in the total
entropy $S$. The transport theoretical description of quarks and
gluons in the PHSD is based on a Dynamical QuasiParticle Model
(DQPM) for partons that is {constructed} to reproduce the lattice
QCD (lQCD) results for a quark-gluon plasma in thermodynamic
equilibrium. The DQPM provides the mean-fields for gluons/quarks and
their effective 2-body interactions that are implemented in the
PHSD. For details about the DQPM model and the off-shell
transport{,}  we refer the reader to the review {in}
Ref.~\cite{Cassing:2008nn}.

We stress that a non-vanishing width in the partonic spectral
functions is the main difference between the DQPM and conventional
quasiparticle models~\cite{qp1}. Its influence on the collision
dynamics {can be} seen in the correlation functions, {which,} in the
stationary limit {involve} the off-diagonal elements of the
energy-momentum tensor $T^{kl}$ {and thus give rise to} the shear
viscosity $\eta$ of the medium~\cite{Peshier:2005pp}.  {A} sizable
width is {then essential for} obtain{ing} a small ratio of the shear
viscosity to entropy density $\eta/s$, which results in a roughly
hydrodynamical evolution of the partonic system in PHSD
\cite{Cass08}. The two-particle correlations {resulting from the
finite width of the parton spectral functions} are taken into
account dynamically {in the PHSD} by means of the {\em generalized}
off-shell transport equations~\cite{Cass_off1} that go beyond the
mean field or Boltzmann
approximation~\cite{Cassing:2008nn,Linnyk:2011ee}.

We recall that the PHSD approach has been tested from low SPS to top
RHIC energies {against the measured} rapidity spectra of various
particle species, transverse mass distributions, differential
elliptic flow of charged hadrons as well as {their} quark-number
scaling~\cite{PHSDqscaling}. More recently, higher harmonics in the
azimuthal distribution of charged hadrons in the plane perpendicular
to the beam direction has {also} been
examined~\cite{PHSDasymmetries}. The description of the various data
sets has been found to be surprisingly good for all bulk
observables.

\begin{figure}
\includegraphics[width=0.5\textwidth]{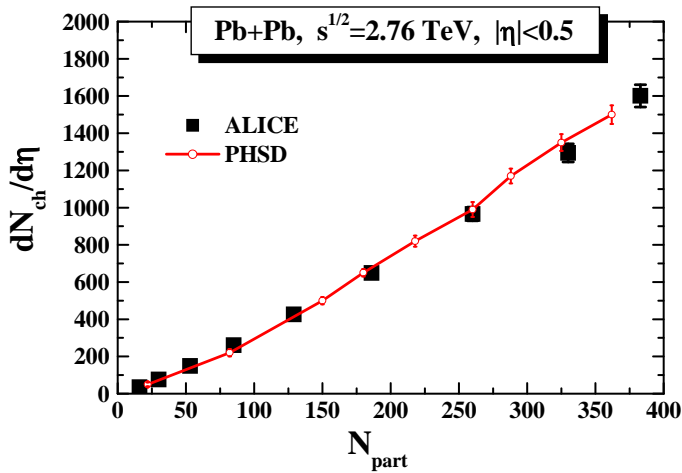}
\caption{(Color on-line) {P}seudo-rapidity distribution of charged
{hadrons} at midrapidity as a function of the number of participants
$N_{part}$ from PHSD (solid line) in comparison to the data from the
ALICE Collaboration \protect\cite{ALICE1} for Pb+Pb at
$\sqrt{s_\mathrm{NN}}$= 2.76 TeV. \label{Fig1} }
\end{figure}

{For} nucleus-nucleus collisions at LHC energies{, one might wonder
if} some new features{, which} are not properly described by the
PHSD approach within the standard settings {of} Ref. \cite{BrCa11}{,
could appear} since in the initial state a color glass condensate
might {become important} \cite{Larry} or the nuclear shadowing could
be different from the extrapolations {given} in Ref.
\cite{BrCa11}{,} etc. To shed some light on these questions{,} we
{have} test{ed} the default PHSD version {by using it to study}
Pb+Pb collisions at $\sqrt{s_\mathrm{NN}}= 2.76$~TeV {and compared
the results} to data from the ALICE Collaboration \cite{ALICE1}. {In
particular, we have examined} the pseudo-rapidity distribution of
charged particles as a function of centrality. In Fig.~\ref{Fig1} we
display the results for $dN_c/d\eta$ at midrapidity from the default
PHSD calculations in comparison to the ALICE data as a function of
the number of participants $N_{part}$ that has been determined
dynamically in the PHSD calculations. {Q}uite acceptable agreement
is seen, {indicating} that the bulk parton dynamics is not much
different at top RHIC and LHC energies. A similar observation has
also been made in Ref.~\cite{Petersen} where the UrQMD transport
approach (without explicit partonic degrees of freedom) was shown to
give a reasonable reproduction of this observable as a function of
centrality. Thus we continue with electromagnetic probes using the
default PHSD version of Ref.~\cite{BrCa11} without readjusting any
PYTHIA parameters {used in the PHSD} for the initial state hard
scattering processes.

\subsection{Partonic sources of dileptons in PHSD}

The PHSD approach so far  has been employed for dilepton production
from $pp$ to Au + Au collisions at SPS \cite{Linnyk:2011vx} and RHIC
energies \cite{Linnyk:2011hz}. As in Refs.
\cite{Linnyk:2011vx,Linnyk:2011hz} dilepton radiation {from} the
constituents of the strongly interacting QGP proceeds via the
following elementary processes: the basic Born $q+\bar q$
annihilation mechanism, gluon Compton scattering ($q+g\to
\gamma^*+q$ and $\bar q+g\to \gamma^*+\bar q$){,} and quark {and}
anti-quark annihilation with {the} gluon Bremsstrahlung in the final
state ($q+\bar q\to g+\gamma^*$). In the on-shell approximation, one
would use perturbative QCD cross sections for the processes listed
above. However, in the strongly interacting QGP the gluon and quark
propagators differ significantly from the non-interacting
propagators. Accordingly, we have calculated in
Refs.~\cite{Linnyk:2004mt,olena2010} the off-shell cross sections
for dilepton production in the partonic channels by off-shell
partons, using the phenomenological parametrizations from the DQPM
for the quark and gluon propagators and their interaction strength.

We have implemented the cross sections obtained in
Refs.~\cite{Linnyk:2004mt,olena2010} into the PHSD transport
approach in the following way \cite{Linnyk:2011hz}: Whenever
quark-antiquark, quark-gluon and antiquark-gluon collisions occur in
the course of the Monte-Carlo simulation of the partonic phase in
{the} PHSD, a dilepton pair can be produced according to the
off-shell cross sections~\cite{olena2010}{. In} addition to the
virtualities of the partons involved, {the latter also depends} on
the energy density in the local cell, where the collision takes
place{, as according to the DQPM it} governs the widths of the quark
and gluon spectral functions as well as the strong coupling.

\subsection{Hadronic sources of dileptons in PHSD}

In the hadronic sector, {the} PHSD model is equivalent to the HSD
transport approach \cite{Cass99,Brat97,Ehehalt}. The implementation
of the hadronic decays into dileptons ($\pi$-, $\eta$-, $\eta '$-,
$\omega$-, $\Delta$-, $a_1$-Dalitz, $\rho\to l^+l^-$, $\omega\to
l^+l^-$, {and} $\phi\to l^+l^-$) in HSD (and PHSD) is described in
detail in Refs.~\cite{Brat08,Bratkovskaya:2008bf}. In contrast to
the HSD approach -- without  explicit partonic degrees of freedom --
the mesons in the PHSD are produced through the dynamical
hadronization from the partonic state. On the other hand, the
subsequent interactions of the produced hadrons as well as the
evolution of the hadronic 'corona' proceed in the same way as in the
HSD.

The PHSD off-shell transport approach is particularly suitable for
investigating the different scenarios for the modification of vector
mesons in a hot and dense medium. {As in} the HSD model, the PHSD
approach incorporates the {\em off-shell propagation} {of} vector
mesons as described in Ref.~\cite{Cass_off1}. In the off-shell
transport, the hadron spectral functions change dynamically during
the propagation through the medium and evolve towards the on-shell
spectral function{s} in the vacuum.   As demonstrated in
Ref.~\cite{Brat08}, the off-shell dynamics is important for
resonances with a rather long lifetime in {the} vacuum but strongly
decreasing lifetime in the nuclear medium (especially $\omega$ and
$\phi$ mesons) and also proves vital for the correct description of
dilepton decays  of $\rho$ mesons with masses close to the two pion
decay threshold. For a detailed description of the off-shell
dynamics and the implementation of vector-meson modifications in the
medium{,} we refer the reader to
Refs.~\cite{Cass_off1,Brat08,Bratkovskaya:2008bf,Linnyk:2011ee,Linnyk:2011hz}.

In Ref.~\cite{Linnyk:2011hz}, the {PHSD has been extended to
include} hadronic sources for dilepton production from secondary
multi-meson interactions {through} the channels $\pi \omega \to
l^+l^-$, $\pi a_1\to l^+l^-$, {and} $\rho \rho \to l^+l^-$. These
so-called `4$\pi$ channels' for dilepton production are incorporated
in the PHSD on a microscopic level rather than assuming thermal
dilepton production rates and a parametrization for the inverse
reaction $\mu^+ + \mu^- \rightarrow 4 \pi's$ is incorporated by
employing the detailed balance as in
{Refs.}~\cite{RH:2008lp,Santini:2011zw}. By studying the
electromagnetic emissivity (in the dilepton channel) of the hot
hadron gas, it was shown in Refs.~\cite{Song:1994zs,Gale:1993zj}
that the dominating hadronic reactions contributing to the dilepton
yield at the invariant masses above the $\phi$ peak are the two-body
reactions of $\pi+\rho$, $\pi+\omega$, $\rho+\rho$, {and} $\pi+a_1$.
This conclusion was supported by the subsequent study in a hadronic
relativistic transport model~\cite{GLi}. Therefore, we implemented
the above listed two-meson dilepton production channels in the PHSD
approach in Ref.~\cite{Linnyk:2011hz}. In addition, some higher
vector mesons ($\rho^\prime$ {\it etc.}) were tacitly included by
using phenomenological form factors {that are} adjusted to {the
experimental} data. {Specifically, we} determined the cross sections
for the mesonic interactions with dileptons in the final state using
an effective Lagrangian approach following the works of
Refs.~\cite{Song:1994zs,GLi}. In order to fix the form factors in
the cross sections for dilepton production by the interactions of
$\pi+\rho$, $\pi +\omega$, $\rho+\rho$ and $\pi a_1$, we used the
measurements in the detailed-balance related channels: $e^+e^-\to
\pi+\rho$, $e^+e^-\to \pi +\omega$, $e^+e^-\to \rho+\rho$, and
$e^+e^-\to \pi+a_1$. Note that we fitted the form factors while
taking into account the widths of the $\rho$ and $a_1$ mesons in the
final state by convoluting the cross sections with the (vacuum)
spectral functions of these mesons (using the parametrizations of
the spectral functions as implemented in the PHSD). In Fig.~5 of
Ref.~\cite{Linnyk:2011hz} we presented the resulting cross sections
{that were} implemented in the PHSD. {Contributions of these
channels to the dilepton invariant mass spectrum in} nucleus-nucleus
collisions at top SPS and RHIC energies {and comparisons to
available data} have been reported in
Refs.~\cite{Linnyk:2011vx,Linnyk:2011hz}.

\section{The extended statistical hadronization model (SHM)}
\label{section_shm}

Dilepton production has {also} been studied in the framework of the
Statistical Hadronization Model (SHM) at
SPS~\cite{Agakichiev:2005ai} and RHIC
energies~\cite{Manninen:2010yf}. Even in the absence of a dynamical
evolution of the fireball, both analys{e}s find fair agreement{s}
with measurements{. In the present work, we}
 extend the previous studies to LHC
energies in order to model in detail the contribution of
 {dileptons} stemming from the so-called
``freeze-out cocktail''. These results can {then} be compared with
our transport simulations in order to quantify the effects from
partonic degrees of freedom {and} the dynamical evolution of the
{produced hot and dense matter}.

{The} SHM is a suitable reference model {for} obtain{ing} additional
dynamical information {about the collisions besides} the particle
spectra at chemical freeze-out. A corresponding comparison for
dilepton spectra from $Au+Au$ collisions at the top RHIC energy {was
presented} in Ref.~\cite{Linnyk:2011hz}. Here we aim at a similar
comparison for $Pb+Pb$ collisions at LHC energies which requires to
specify the intrinsic parameters of the statistical hadronization
model. {Since the} central rapidity {region} is, to a good
approximation, net charge free {in heavy ion collisions} at {the}
LHC energies{,} we set all chemical potentials to zero in this
analysis. Based on findings at SPS and RHIC, we expect the
$\gamma_S$ (strangeness suppression) factor to be close to unity in
central $Pb+Pb$ collisions at LHC energies. This was recently
confirmed in {Ref.}~\cite{Preghenella:2012eu} where measurements of
ratios of hadronic rapidity densities were compared with statistical
model predictions~\cite{Andronic:2008gu} evaluated with the thermal
parameter $T=164$MeV and $\gamma_S$=1. All ratios of kaons and
multi-strange hyperons to pions were correctly predicted{,}
indicating that strangeness is indeed chemically equilibrated (i.e.
$\gamma_S=1$) in central $Pb+Pb$ collisions at {the} LHC energies.
Therefore, the thermal state of the {produced} fireball can be
specified with two free parameters: the temperature $T$ and volume
$V$.
 We do not attempt to extract the chemical freeze-out
temperature from the data but use the predicted asymptotic value of
$T$=170 MeV~\cite{Manninen:2008mg} instead. The overall
normalization is determined from the measured transverse momentum
spectra of various hadrons (see below).

The SHM - describing a static single fireball - is rather reliable
in predicting the phase-space integrated relative yields of
different hadrons. However, since experiments measure electrons and
positrons at mid-rapidity only and above some minimum transverse
momentum, we need to emulate the initial state dynamics such that
the detector acceptance is correctly accounted for.

The ALICE collaboration has measured the transverse momentum spectrum of
$\pi^-$, $K^-$ and $\bar{p}$ in the 0-5\% most central Pb+Pb
collisions at 2.76 TeV~\cite{Preghenella:2012eu} while $\Xi^-$ and
$\Omega^-$ spectra are measured in the 0-20\% most central
collisions. We can estimate the multi-strange hyperon rapidity
densities in the 0-5\% most central collisions with help of the
combined ALICE and CMS measurement of charged hadron rapidity
density as a function of centrality (see Figure 5 of
~\cite{Chatrchyan:2011pb}).

{Fitting} the above-mentioned charged hadron data {by a second-order
polynomial,} { we obtain the following} scaling factor{:}
\begin{equation}
N_{5 \rightarrow 20\%} = \frac{\d N_{ch}^{0-5\%}}{\d y} / \frac{\d
N_{ch}^{0-20\%}}{\d y} = 1612/1203 = 1.34{,}
\end{equation}
by which we multiply the hyperon transverse momentum spectra in
{the} 0-20\% most central collisions in order to estimate the
corresponding spectra in the 0-5\% most central collisions.

{We then} assume that the hadron {yield} in $Pb+Pb$ collisions
arise{s} as a superposition of \NPt$/2$~independent nucleon-nucleon
collisions and determine the average elementary volume
$V_{pp}=2{.}70$ fm$^3$ such that our pion yield, which scale like
$\NP$$V_{pp}/2$, agree with the measured pion yield in central
$Pb+Pb$ collisions. We {further take} \NPt=383~independent
nucleon-nucleon collisions for each 0-5\% most central $Pb+Pb$
collision.

{The total} mass $M_{\rm clust}$ of the {hadron} cluster {in the
fireball can be calculated} by {adding} the energies of individual
hadrons. {From} the equipartition theorem {for a system in thermal
equilibrium that} energy is shared equally among its {degrees of
freedom,} we can assume that the ``thermal'' momentum of the cluster
$|\vec{p}_{\rm clust}|$ equals on average {to} the mass of the
cluster. {Since the longitudinal direction is dominantly governed by
the initial state parton dynamics, we only apply this to the
cluster's transverse momentum, i.e.,} {its average transverse
momentum is} $\langle p_T^{\rm
clust}\rangle=\frac{\sqrt{2}}{\sqrt{3}}M_{\rm clust}$. In each
event, we {thus} first calculate $M_{\rm clust}$ and sample a
Gaussian distribution for the cluster{'}s {transverse momentum} such
that the mean value of the {G}aussian equals {to}
${\sqrt{2/3}}M_{\rm clust}$ while the width of the distribution is
taken as $M_{\rm clust}{/\sqrt{6}}$.

{For the distribution of the longitudinal momentum of the cluster,
we} rely here on {the} simple Landau scaling, which is in fair
agreement with the measurements at least up to {the} RHIC energies,
for a discussion, see e.g. {Ref.}~\cite{Bleicher:2005ys}. According
to the Landau scaling, the width of the rapidity distribution of
pions can be estimated from the simple formula
\begin{equation}
\sigma_y^{\pi}(\sqrt{s})=\ln( \frac{\sqrt{s}}{2m_p}) \,\,\,\,\,\,\,
\,\,\,\,\,\,\, \,\,\,\,\,\,\, m_p=0.94 {\rm GeV},
\end{equation}
leading to $\sigma_y^{\pi}$(2.76TeV)=7.26 which we take as the width
of the clusters' rapidity distribution in our calculations. We note
 that in our previous work at {the} RHIC
{energies,} we have fitted the width of the clusters' rapidity
distribution {with} the value $\sigma_y^{\pi}$(200GeV)=4.2 which is
in fair agreement with the {value} $\sigma_y^{\pi}$(200GeV)=4.7
{obtained from the Landau scaling}. {All} hadrons {are then boosted}
to the laboratory frame where the cluster moves with the momentum
$\vec{p}_{clust}$= ($p_T^{\rm clust},y_{\rm clust}\sqrt{p_T^{2\rm
clust}+ M^2_{\rm clust}})$.

\begin{figure}
\includegraphics[width=0.46\textwidth]{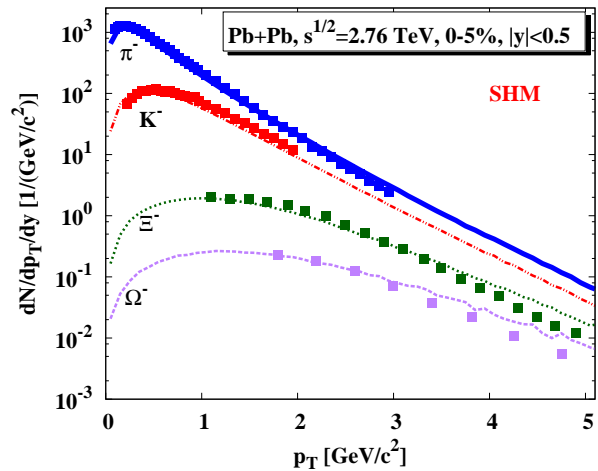}
\caption{(Color on-line)\label{SHMpTspectra} Transverse momentum
spectra of $\pi^-$, $K^-$, $\Xi^-$ and $\Omega^-$ (squares) in
$Pb+Pb$ collisions at 2.76 TeV~\cite{Preghenella:2012eu} measured
by the ALICE collaboration in comparison to our statistical model calculations
(lines).}
\end{figure}

A comparison of our {calculated} transverse momentum spectra of
$\pi^-$, $K^-$, $\Xi^-$ and $\Omega^-$ in central $Pb+Pb$ collisions
with ALICE measurements is presented in Figure.~\ref{SHMpTspectra}.
We {have omitted here} the transverse momentum spectrum of $\bar{p}$
{as it} is not in agreement with {that measured in experiments}.
Nevertheless, the midrapidity data for the different hadron species
is rather well described such that we can continue with the {study
of} hadronic dilepton decays.

The form factors for dilepton decays of the light hadrons employed
here are the same as in our previous study~\cite{Manninen:2010yf}.
Since the low-mass dilepton spectrum is strongly dominated by
hadronic decay channels, the SHM already allows {us} to evaluate the
$e +e^-$ pair spectrum from the different hadronic decays, i.e. the
usual 'hadronic cocktail'. On the other hand, above invariant masses
of about 1 GeV the dominant hadronic channels {for} dileptons are
the correlated and un-correlated semi-leptonic decays of open charm
and open beauty mesons, whose {\em relative populations} are modeled
within the extended SHM while the total open charm and beauty cross
sections are taken from elementary p+p collisions at the same
bombarding energy and scaled with the number of binary
nucleon-nucleon collisions (see below).

The uncertainty in the total charm and beauty production cross
section{s}, especially at LHC energies, is sizable. In order to
estimate these cross sections{,} we use different (but related)
models to achieve a good reproduction of {the} data {from} $pp$
collisions at \sqrtsnn =2.76 TeV {as described} in the following
Section~\ref{sectioncharm}. The resulting cross sections {are} used
to {calculate} the production of $D$- and $B$-mesons in $Pb+Pb$
collisions at \sqrtsnn =2.76 TeV{. Contributions from leptonic
decays of these heavy mesons are then added} to {the dilepton
spectra obtained from} SHM and PHSD {described} in
Section~\ref{sectionresults}.

\section{Open charm and beauty}
\label{sectioncharm}

\subsection{Open charm and beauty in proton-proton collisions}

There are experimental indications~\cite{Andronic:2006ky,baumgart}
that heavy flavor production might also exhibit statistical features
in high energy nuclear as well as in $e^+e^-$ collisions{,} at least
{for the relative abundances of} open charm~\cite{Andronic:2009sv}
and possibly also for {that of} open beauty
mesons~\cite{Becattini:2008tx}. Since heavy-flavor production occurs
early in the reactions and is not {likely} formed thermally, we
{thus extend the} statistical hadronization model {to include} open
charm and beauty
 {mesons} such that the{ir} relative yields are evaluated
with the SHM (for some freeze-out temperature $T$) while the{ir}
total {numbers are estimated from measured} open and beauty
{production} cross sections. {For the $D_s$ and $B_s$ states, their
production cross sections are multiplied by the factor
$\gamma_s^{\rm hard}$=0.3 relative to those for nonstrange heavy
mesons to take into account the strangeness suppression in hard
proton-proton scattering. Furthermore, we take into account only the
6 lowest mass states of open charm mesons in the extended SHM since
they already account for more than 90\% of the total open charm
production cross section and also not much are known about the
higher mass excited states.}

{With the above extended SHM, we} study the open charm and beauty
production for the two freeze-out temperatures $T$=170 MeV and 150
MeV while the chemical potentials are kept zero. The {resulting}
relative normalized primary and final (after strong decays)
production probabilities  for the open charm states (calculated
within the SHM) are given in Table~\ref{Dtable}. {The} ratios of
{final} open charm hadrons {can then be calculated and compared with
those measured by the LHCb and ALICE Collaborations in $p+p$
collisions at \sqrtsnn = 7 TeV}
~\cite{Gersabeck:2010bd,LHCb_openD,ALICE:2011aa} as {shown in} Table
\ref{Dtable2}. As one can see, all ratios are reproduced (within
errors) for both temperatures of 150 and 170 MeV. We choose here, as
for the light hadrons, $T$=170 MeV as our temperature for all hadron
species considered in the analysis. {We note that} the feed-down
{contribution} from beauty mesons {was neglected in the above
analysis} since this contribution is estimated to be of the order of
10\% of the total charm production cross section~\cite{Bala:2012vg}.

\begin{table}
\begin{tabular}{|c|c|c|c|c|}
\hline
Hadron & \multicolumn{2}{|c|}{primary} & \multicolumn{2}{|c|}{final state} \\
\hline
& T=170 MeV & T=150 MeV & T=170 MeV & T=150 MeV \\
\hline
$D^+$    &  0.186 & 0.200 & 0.273 & 0.286 \\
$D^0$    &  0.190 & 0.204 & 0.646 & 0.642 \\
$D_s$    &  0.0333 & 0.0331 & 0.0810 & 0.0751 \\
$D^{*+}$ &  0.269 & 0.258 & 0 & 0 \\
$D^{*0}$ &  0.274 & 0.263 & 0 & 0\\
$D_s^*$  &  0.0474 & 0.0420 & 0 & 0 \\
\hline
\end{tabular}
\caption{Relative production probabilities of open charm mesons
before (primary) and after (final) strong decays.} \label{Dtable}
\end{table}

\begin{table}
\begin{tabular}{|c|c|c|c|c|}
\hline
 & LHCb~\cite{Gersabeck:2010bd,LHCb_openD} & ALICE~\cite{ALICE:2011aa} & T=170 MeV & T=150 MeV \\
\hline
$\frac{D^0}{D^{*+}}$ & 2.20 $\pm$ 0.48 & 2.09 & 2.40 & 2.49 \\
$\frac{D^0}{D^{+}}$  & 2.07 $\pm$ 0.37 & 2.08 &  2.37 & 2.25 \\
$\frac{D^0}{D_s}$   & 7.67 $\pm$ 1.67 & & 7.98 & 8.55 \\
$\frac{D^{*+}}{D^+}$ & 0.94 $\pm$ 0.22 & 1.00 & 0.99 & 0.90 \\
$\frac{D^{*+}}{D_s}$ & 3.48 $\pm$ 0.93 & & 3.32 & 3.44 \\
$\frac{D^+}{D_s}$    & 3.70 $\pm$ 0.84 & & 3.37 & 3.81 \\
\hline
\end{tabular}
\caption{Ratios of
measured~\cite{Gersabeck:2010bd,LHCb_openD,ALICE:2011aa} open charm
yields in $p+p$ collisions at 7 TeV compared with SHM model
calculations for the temperatures $T$=170 MeV and $T$=150 MeV.}
\label{Dtable2}
\end{table}

{In the above, we have taken} the total open charm production cross
section in $p+p$ reactions at 7 TeV to be
$$\sigma_{\bar{c}c}^{\rm 7TeV}=6.4 \mbox{ mb}$$
{from} fitting a Gaussian to the combined measurement{s} of ALICE
and LHCb $D$ {on} meson rapidity distributions. Again, the
longitudinal dynamics of open heavy flavor mesons are evaluated in
the model of Ref.~\cite{Manninen:2010yf} while the transverse
momentum distributions are predicted by the MC@sHQ
model~\cite{Gossiaux:2008jv,Gossiaux:2009hr,Gossiaux:2010yx},
explained in Section~\ref{SectNantes}. Our estimate for the total
charm production cross section (6.4 mb) {is comparable to the}
preliminary LHCb estimate of $6.1\pm0.93$~mb obtained by tuning
PYTHIA to the LHCb measurement and then integrating over the whole
phase-space~\cite{LHCb_openD}. An extrapolation of the ALICE
collaboration measurement at mid-rapidity to full phase-space
yields{, however, a} cross section of $\sigma_{\bar{c}c}^{\rm
7TeV}=8.5^{+4.2}_{-2.4}$~mb~\cite{ALICE:2012sx}. The ATLAS
collaboration gives{, on the other hand,} for the total cross
section a preliminary value of
$7.13^{+4.0}_{-2.2}$~mb~\cite{ATLASCONF}. The three LHC experiments
agree within the uncertainties. Perturbative QCD calculations were
also found to agree with the data and with our estimate for the
total charm production cross section in $p+p$ reactions at 7~TeV
(see~\cite{Carrer:2003ac,ALICE:2011aa} and references therein).

{For} the total charm production cross section at 2.76 TeV reads{,
it is estimated to be} $\sigma_{\bar{c}c}^{\rm 2.76TeV}$=4.8$\pm$0.8
mb {by the ALICE Collaboration}. Using the ratio of the two ALICE
cross sections (at 7 TeV and at 2.76 TeV), we find that our estimate
for the total cross section of charm production in $p+p$ collisions
at the lower LHC energy  {is}
$$\sigma_{\bar{c}c}^{\rm 2.76TeV}=3.6\mbox{ mb{,}}$$
{which is used in our study for heavy ion collisions} at 2.76~TeV.

For the open charm decays, we have implemented in detail the form
factors for each of the semi-leptonic decay channels that are
{inlcuded} in {our} analysis. A discussion and presentation of
relevant formulae for the open charm form factors can be found in
the Appendix~\ref{apex}. For beauty decays, no form factors were
employed.

The properties of open beauty mesons are not well known
experimentally and thus we use the known properties from the
PDG~\cite{Amsler:2008zzb} and take the missing information from
model calculations. In the case of open beauty, we take into account
18 lightest states and their anti-particles. Only the lowest lying
states $B^+$, $B^0$, $B_s$ and $B^*_s$ and their anti-particles
decay semi-leptonically into electrons, while the excited states
decay under the strong interaction into the lower mass states. It is
important to take into account also the excited states in the
analysis in order to properly evaluate the total beauty cross
section and the relative abundances of the lowest lying $B$ meson
states. We ignore in this work the $b$-baryons whose production
cross section is expected to be  small ($\approx$10\% of
$\sigmabb$)~\cite{Amsler:2008zzb} and thus our $\sigmabb$ refers to
the total cross section for the production of open beauty {\em
mesons}.

The properties (masses and decay channels) as well as the references
to the data and models that we use for our open beauty calculations
are listed in Table \ref{beautytable} {in Appendix~\ref{apex}}. In
all cases we have taken a weighted average over the different
estimates as input {to our study}{, and} a recent compilation for
the numerical values of the open beauty masses can be found
in~\cite{Vijande:2007ke}.

Measurements of {the} total beauty cross section is not yet
available at LHC energies, hence we {relate it} to the charm cross
section and use available theoretical estimates for the ratio of
these cross sections to estimate $\sigmabb$ in p+p collisions at
2.76 TeV. We note that perturbative QCD predictions for this ratio
suffer from large uncertainties~\cite{VogtLHCBB}. Using here the
same model and parameters as those applied for the calculation of
the energy loss of heavy quarks in the next Section, we obtain the
ratio $\sigmacc/\sigmabb$=40 in central $Pb+Pb$ collisions at 2.76
TeV.

\subsection{Energy loss of D- and B-mesons in $Pb+Pb$ collisions}
\label{SectNantes}

{The} total cross sections and spectra of D-mesons {obtained in the
previous Section from the available experimental data are for}
elementary p+p collisions. {We are} interest{ed}
 in the
contribution of the decays of correlated and uncorrelated $D+\bar D$
and $B+\bar B$ pairs to the dilepton spectra {in heavy ion
collisions, but simple} binary scaling of the spectra {from} p+p
collisions {is expected to overestimate their} contribution{s}.
{This is} because {of the neglect of} the loss of correlation and
the modification of the spectra of heavy hadron pairs due to the
partonic energy loss~\cite{Gossiaux:2010yx} and hadronic
rescattering~\cite{Linnyk:2008hp}. {To} take into account the
collisional and radiative energy loss of charm and beauty quarks in
central Pb+Pb collisions{, we use} the event generator for heavy
quarks and mesons of P.B.~Gossiaux {\it et al.} -- Monte Carlo for
Heavy Quarks (MC@sHQ). This model -- described in detail in
Refs.~\cite{Gossiaux:2008jv,Gossiaux:2009hr,Gossiaux:2010yx} --
assumes that in heavy-ion collisions the initial momentum
distribution of charm and bottom quarks is identical to that in p+p
collisions which are predicted by {the} FONLL {PQCD} calculations
\cite{Cacciari:2005rk} and verified experimentally by the STAR and
PHENIX collaboration{s} \cite{Adamczyk:2012af}. {With the} expansion
of the {produced quark-gluon} plasma
 described by {the hydrodynamic
model, the} interaction of {heavy quarks with} the plasma
constituents is {then} calculated {from the PQCD} one-gluon-exchange
{diagrams} {but with} a running coupling constant and an infrared
regulator adjusted to reproduce the same energy loss as {in the}
hard thermal loop calculations. Both collisional and radiative
energy loss are taken into account {and with the} latter corrected
by the Landau-Pomeranschuk-Migdal effect. This model {has been shown
to} describe {very well} the nuclear modification factor $R_{AA}$
and the elliptic flow $v_2$ of heavy mesons (non{-}photonic
electrons) measured at RHIC and more recently {at LHC} by the ALICE
Collaboration~\cite{Aichelin:2012ww}.

Here, we employ the transverse momentum distributions of the $D$-
and $B$-mesons produced in central Pb+Pb collisions as calculated in
the MC@sHQ event generator as an input for our calculation of the
dilepton yield by the semi-leptonic decays of $D+\bar D$ and $B+\bar
B$ meson pairs. In particular, the decay of a single $D$ ($\bar D$)
meson is simulated and the produced electron (positron) is combined
with the positron (electron) produced in the decay of the $\bar D$
($D$) meson coming from the same production vertex to form dilepton
pairs from the {\em correlated} $D+\bar D$ decays. A $D$-meson-decay
electron is combined with a positron produced in the decay of a
$\bar D$-meson coming from a (random) different production vertex
but from the same event to form dilepton pairs from the {\em
uncorrelated} $D+\bar D$ decays. Furthermore, our prescription for
estimating the loss of correlations due to the interactions of charm
in the fireball is to multiply the yield from the correlated pairs
by the suppression factor $R_{AA}^2$, where $R_{AA}$ is the measured
nuclear modification factor for the $D$-mesons. {A similar}
procedure {is used} for {treating the semi-lepton decays of}
$B$-mesons.

\section{Charmonia}
\label{section_cc}

\subsection{Charmonia in proton-proton collisions}

\begin{figure}
\includegraphics[width=0.46\textwidth]{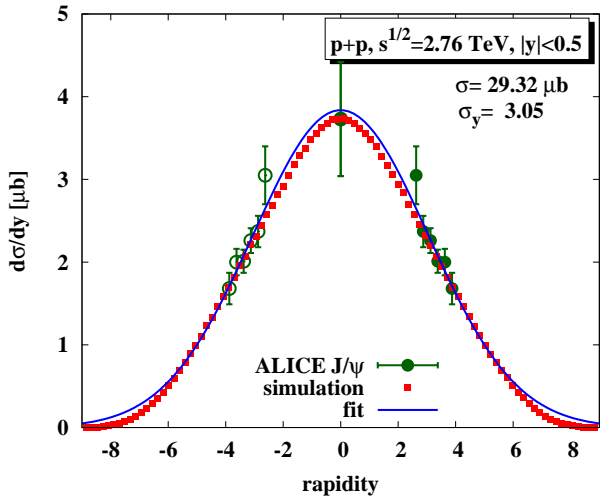}
\caption{(Color on-line)\label{jpsirapidity} \jpsi~rapidity
distribution in p+p collisions at 2.76 TeV measured by the ALICE
Collaboration~\cite{Abelev:2012kr} (full spheres) and reflection of
the data (open spheres). {S}olid line {is a Gaussian fit to the data
based on a} total \jpsi~production cross section of 29.3$\mu b$ for
this collision system. Also shown  {by} squares {is} our simulated
\jpsi~rapidity distribution as described in the text.}
\end{figure}

{Quarkonia are important (and experimentally clean) sources of
dilepton in the high mass region and can also be used to constrain
and cross-check our results regarding open charm production.
Unfortunately, the SHM is not a suitable model for studying the
production of the hidden charm hadrons. Therefore, we} include the
\jpsi~and $\psi$' mesons in our analysis as {an} external input
{taken} from the experiments, i.e. we do minimal modeling for these
hadrons and take basically all necessary information from available
data.

The ALICE collaboration has measured the \jpsi~rapidity and
transverse momentum distributions in $p+p$ collisions at 2.76
TeV~\cite{Abelev:2012kr}. We fit a Gaussian to the measured rapidity
distribution in order to extract the inclusive \jpsi~production
cross section of 29 $\mu b$ for this collision system {as shown by
the solid line in} Fig.~\ref{jpsirapidity}. According to the color
evaporation model, the relative production probabilities of
different quarkonia are beam-energy independent at
ultra-relativistic energies (for a recent review, see
{Ref.}~\cite{Kluberg:2009wc}) and, once the \jpsi~cross section is
known, we can estimate the production cross section of $\psi$' in
$p+p$ collisions at 2.76 TeV with the help of the world
average~\cite{Faccioli:2008ir} $\frac{\psi ' \rightarrow
J/\psi}{J/\psi}$=8.1\%.

The transverse momentum {spectrum} of \jpsi {can} also {be}
determined from the ALICE data~\cite{Abelev:2012kr}{,} and we use
the same {spectrum} for \jpsi~and $\psi$' both in $p+p$ as well as
in $Pb+Pb$ collisions. The parameters characterizing the transverse
momentum spectra of hidden and open heavy flavor hadrons are
collected in Table~\ref{Bntable} {in Appendix~\ref{apexb}}. We
mention here that the details of the $p_T$ spectrum play only a
minor role in di-electron radiation from the quarkonia, while in the
case of $D$ and $B$ meson decays the results are much more sensitive
to the transverse momentum spectrum. Due to this reason, we can use
the same $p_T$ profile for quarkonia in $p+p$ and $A+A$ while for
open heavy flavor decays the {two} systems ha{ve} to be modeled
separately.

The longitudinal momentum distributions of all heavy flavor mesons
are modeled phenomenologically as explained in detail in our earlier
work~\cite{Manninen:2010yf}. In the present calculations, we {use
the same} relevant parameters (the choice of parton distributions,
which determines the width of the clusters' distribution in
longitudinal direction) which we have fixed at RHIC energies.
{Furthermore, we} use the CT10~\cite{Lai:2010vv} parametrization for
{the} parton distributions in all calculations.  The calculated
rapidity distribution of \jpsi's within this model (with the
pre-determined cross section) is compared to the {ALICE} data and to
the Gaussian parameterizations in Fig.~\ref{jpsirapidity}. We find
that the {measured} \jpsi~rapidity distribution is nicely reproduced
and will use the same procedure to evaluate the longitudinal
{momentum distributions} of all heavy flavor hadrons in this work.
We mention here (without explicit presentation) that the {measured}
rapidity distributions of open charm hadrons {at both}
mid-rapidity~\cite{ALICE:2011aa} and forward
rapidities~\cite{Gersabeck:2010bd,LHCb_openD} {in $p+p$ collisions
at 7 TeV} are also in good agreement with {the results from} this
model.

\subsection{Nuclear modification of charmonia}
\label{SectTAMU}

To estimate the $J/\Psi$ contribution {to} the dilepton spectrum {in
heavy ion collisions}, we need information on the $J/\psi$ nuclear
modification factor in {these} collisions. This can be determined
using the approach of Refs.~\cite{Song:2010er,Song:2011xi} that
includes charmonium production from both initial hard
nucleon-nucleon scattering and regeneration from charm and anticharm
quarks in the produced QGP. For the initially produced $J/\Psi$,
their number is proportional to the number of binary collisions
between nucleons in the two colliding nuclei. Whether these
$J/\Psi$'s can survive after the collision depends on many effects
from both the initial cold nuclear matter and the final hot partonic
and hadronic matters. The cold nuclear matter effects include the
Cronin effect of gluon-nucleon scattering before the production of
the primordial $J/\Psi$ from the gluon-gluon
fusion~\cite{Cronin:1974zm}, the shadowing effect due to the
modification of the gluon distribution in a heavy
nucleus~\cite{Eskola:2009uj} and the nuclear absorption by the
passing
nucleons~\cite{Alessandro:2003pi,Lourenco:2008sk,Vogt:2010aa}.  For
the hot partonic and hadronic matter effects, they include the
dissociation of charmonia in the QGP at temperatures higher than the
dissociation temperature and the thermal decay of surviving
charmonia through interactions with thermal partons in the expanding
{QGP}.  In studying these effects, medium effects on the properties
of the charmonia and their dissociation cross sections are further
taken into account by using the screened Cornell potential
model~\cite{Karsch:1987pv} and the NLO pQCD~\cite{Park:2007zza}.
This leads to a dissociation temperature of $\sim 300$ MeV for
$J/\psi$ and $\sim T_C=175$ MeV for $\Psi^\prime$, where $T_C$ is
the QCD phase transition temperature. Although the $J/\Psi$ can
survive above $T_C$, its thermal decay width is not small with a
value of $\sim 10$ MeV at $T=200$ MeV and increasing to $\sim 100$
MeV at $T=280$ MeV. Because of the appreciable number of charm
quarks produced in relativistic heavy-ion collisions, charmonia can
also be regenerated from charm and anticharm quarks in the QGP. A
rate equation is then used to include the effect of thermal
dissociation and regeneration of charmonia. Since charm quarks are
not expected to be completely thermalized -- neither chemically nor
kinetically -- during the expansion of the hot dense matter, a
fugacity parameter and a relaxation factor are introduced to
describe their distributions. Modeling the evolution of the hot
dense matter produced in relativistic heavy-ion collisions by a
schematic viscous hydrodynamics~\cite{Song:2010er}, the average
temperature of the initially produced QGP in central Pb+Pb
collisions at $\sqrt{s_{NN}}=2.76$ TeV is $\sim 311$ MeV if the
initial thermalization time is taken to be 1.05 fm/$c$. The
resulting nuclear modification factors for $J/\Psi$(1S) and
$\Psi^\prime$(2S) are 0.30 and 0.39, respectively, if the shadowing
effect is included. These values increase to 0.57 and 0.95,
respectively, in the absence of the shadowing effect.

\section{Results}
\label{sectionresults}

\subsection{The baseline: p+p collisions}

\begin{figure}
\centering
\includegraphics[width=0.5\textwidth]{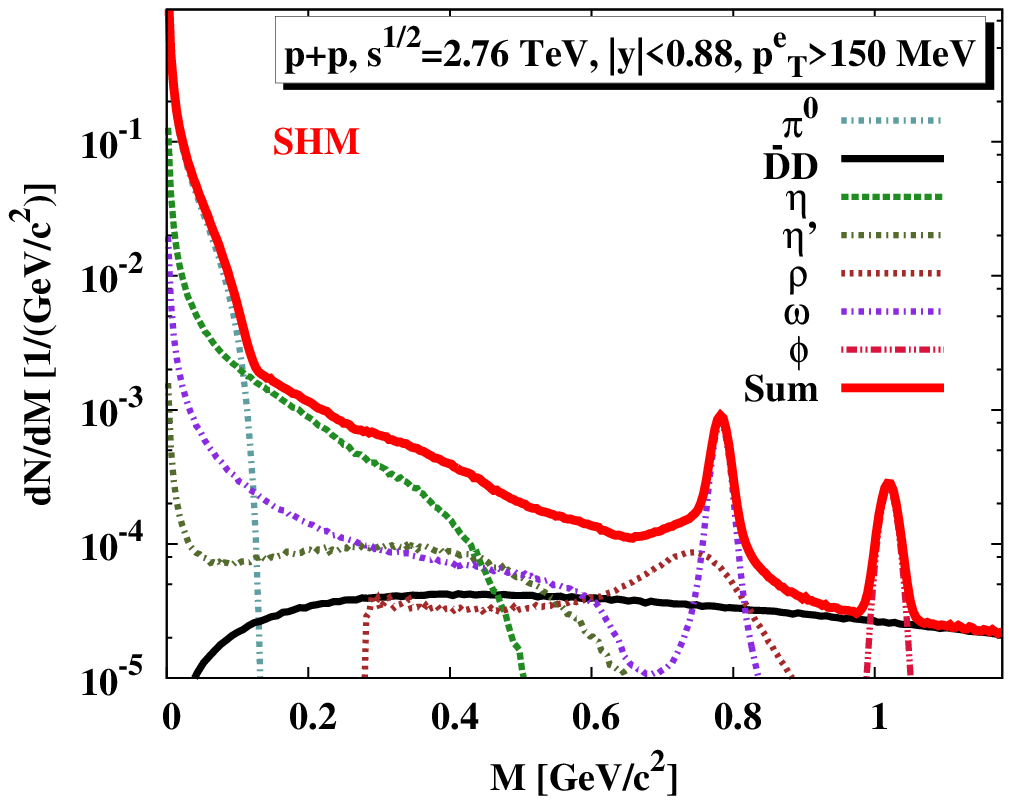}
\includegraphics[width=0.5\textwidth]{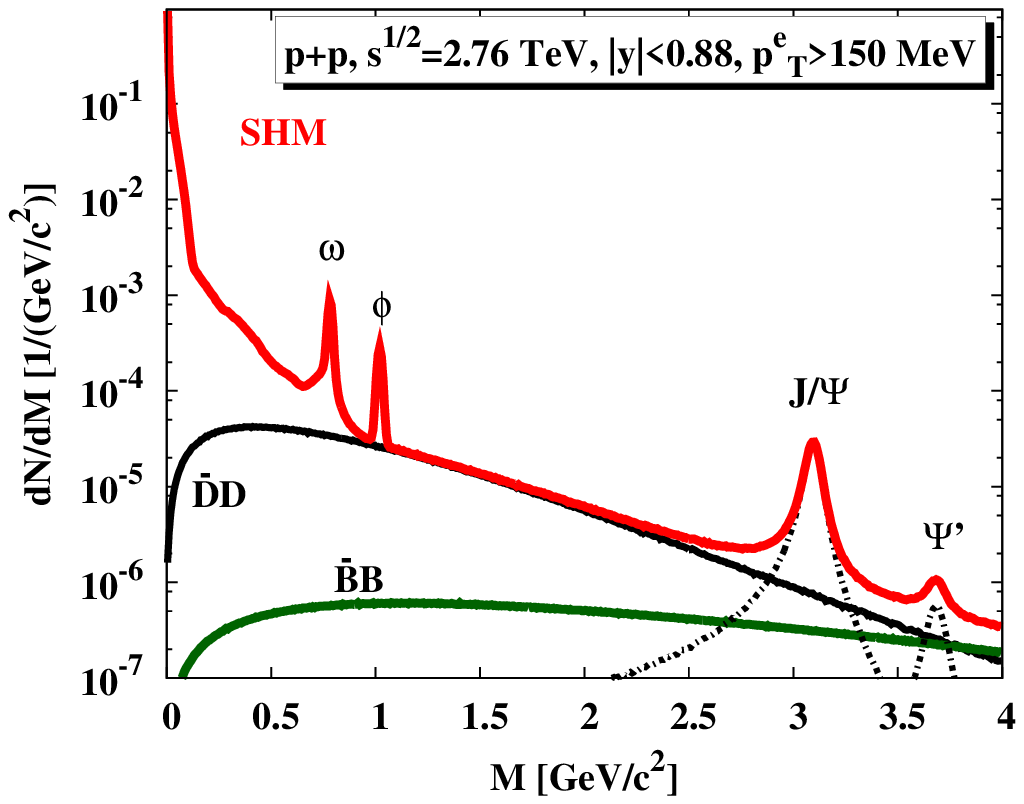}
\caption{(Color on-line) Dielectron invariant mass spectrum from
various different sources evaluated for $p+p$ collisions at 2.76 TeV
within the SHM {for the invariant} mass range up to {1.2 GeV (upper
window) and} 4 GeV {(lower window)}.} \label{Jaakko276}
\end{figure}

\begin{figure}
\centering
\includegraphics[width=0.5\textwidth]{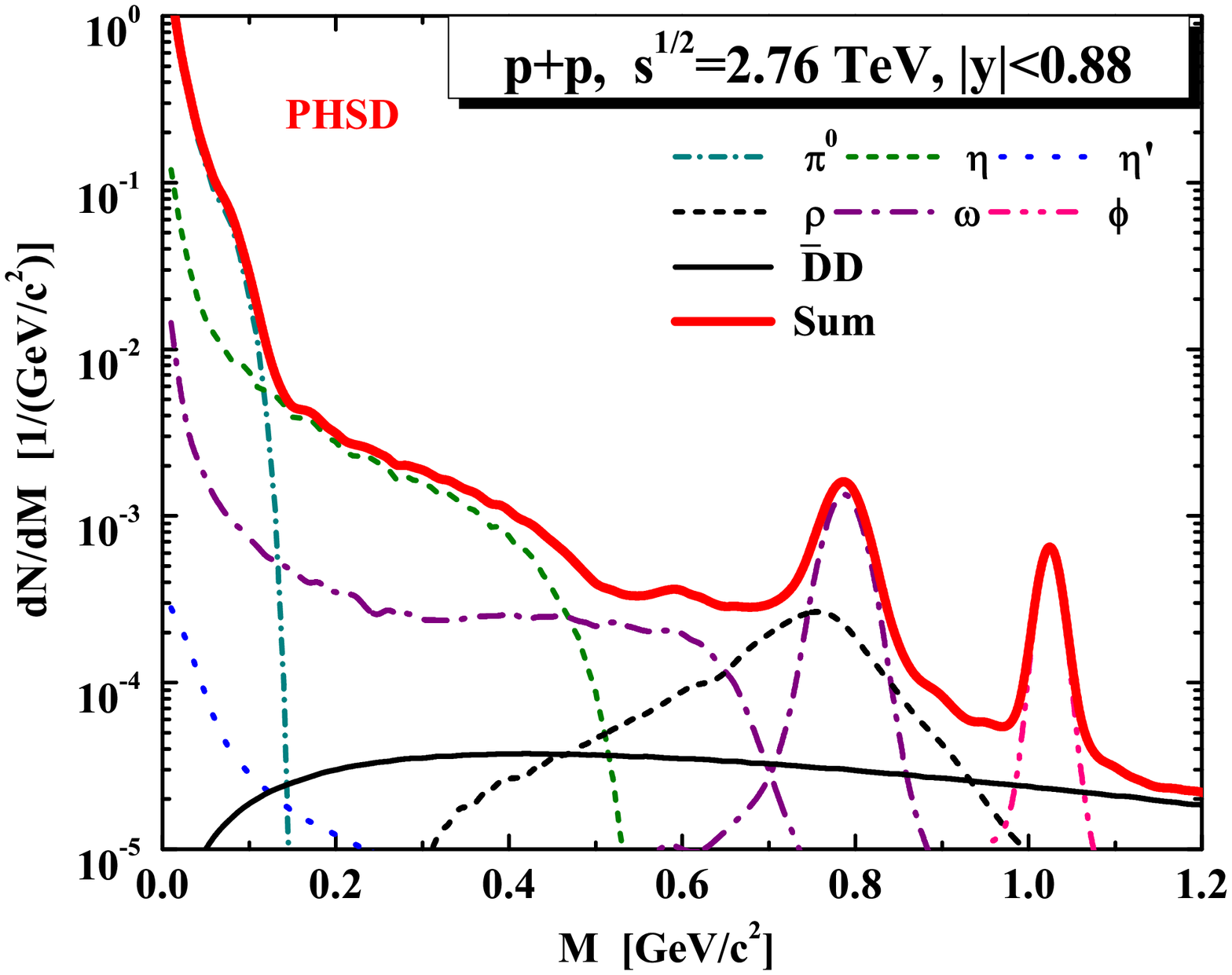}
\includegraphics[width=0.5\textwidth]{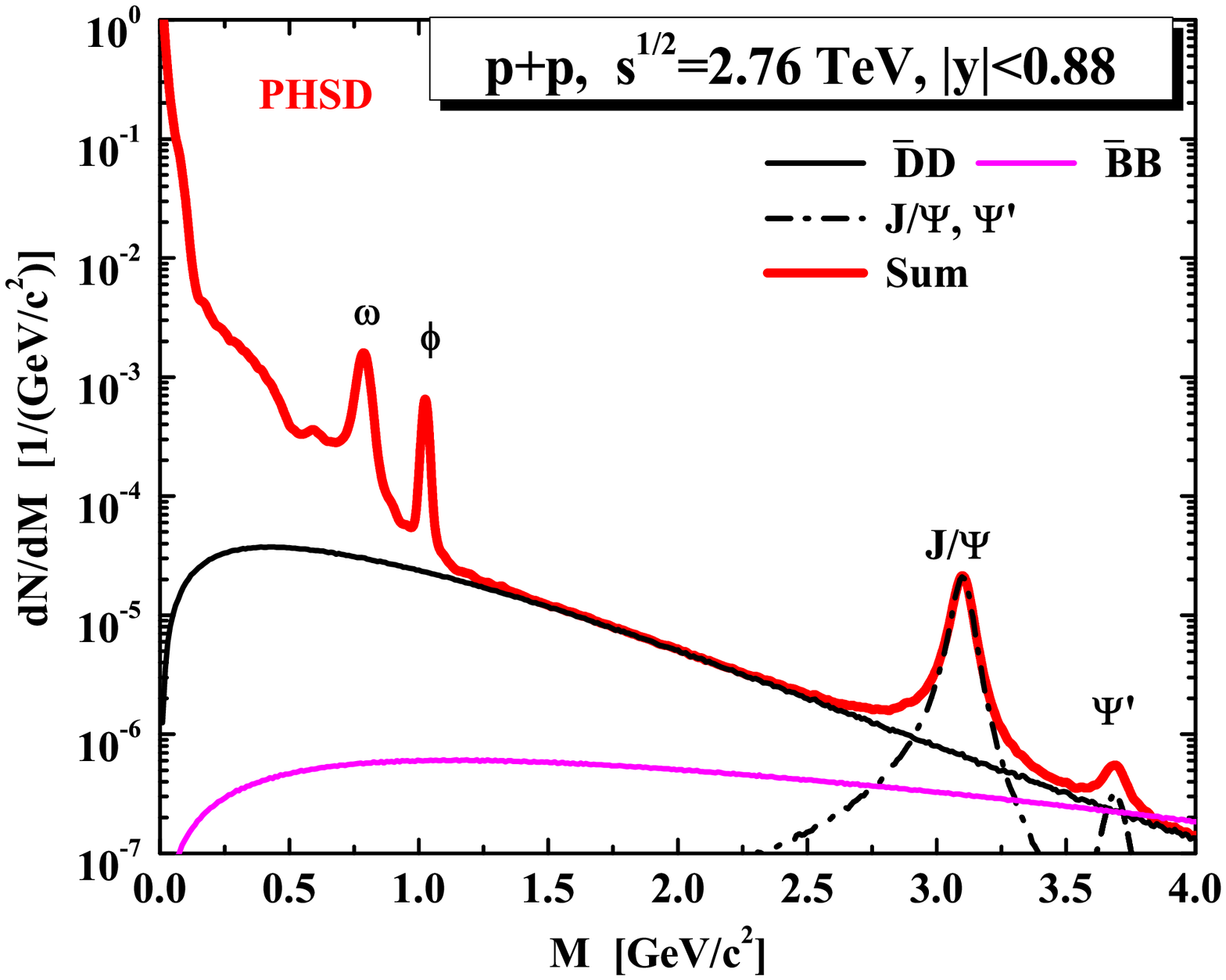}
\caption{(Color on-line) {Same as Fig.\ref{Jaakko276} from the PHSD
approach.}} \label{PHSDpp}
\end{figure}

As a baseline {for our dilepton studies}, we first show in
Fig.~\ref{Jaakko276} the SHM results for the invariant mass spectrum
of $e^+ e^-$ pairs produced in $p+p$ collisions at \sqrtsnn = 2.76
TeV. Qualitative{ly}, the spectra are very similar to {those from}
the top RHIC energy of \sqrtsnn = 200~GeV~\cite{Manninen:2010yf} but
enhanced by roughly a factor of $2$. {These results have been
obtained with} a transverse momentum cut of 150~MeV and a rapidity
cut of $|y_e|<0.88$ for the single electrons, {similar to} the
nominal acceptance cuts of the ALICE detector~\cite{Aamodt:2010jd}.
{Also,} the centrality and $p_T$ dependent parameterizations of the
ALICE detector resolution ~\cite{Aamodt:2010jd} ha{v}e been {used
for determining the mass resolution}. We have {further included} the
contribution from the initial state Drell-Yan process in the
collinear factorization approach of the next-to-leading-order
perturbative
QCD~\cite{Altarelli:1978id,Altarelli:1979ub,KubarAndre:1978uy} (see
also {Ref.}~\cite{GolecBiernat:2010de} for a recent discussion at
LHC energies), which is expected to be compatible with {the
experimental} data within a factor of two. {We find that the}
Drell-Yan process contribute{s} less than 1\% in the {dilepton
invariant mass} range {considered in} our studies and thus discard
its contribution.

Next, we present in Fig.~\ref{PHSDpp} the PHSD results for the
invariant mass spectrum of $e^+ e^-$ pairs produced in $p+p$
collisions at \sqrtsnn = 2.76 TeV, employing the same transverse
momentum and rapidity cuts for the single electrons, i.e.
$p_T^{e}>150$ MeV and $|y_e|<0.88$. We observe generally a good
agreement between the two models.  The minor differences in the line
shapes stem from i) the production of hadrons in proton-proton
collisions in the PHSD from string decays versus the thermal
distribution of primary produced particles (combined with the
feed-down from resonances) in the SHM and ii) the slightly different
compositions of dilepton-decay channels taken into account in the
SHM and the PHSD. The list of the channels for the meson decays to
 dileptons - contributing to the di-electron spectrum in
proton-proton collisions - is presented in
Table~\ref{dilepton_channels_list}. The (non-dominant) channels,
which are taken into account in the SHM but not in the PHSD, are
marked with asterisks.

\begin{table}
\begin{tabular}{|c|c|c|c|}
\hline
 & direct & Dalitz & other \\
\hline
$\pi^0$ & - & $\pi^0\rightarrow \gamma\, \e^+\e^-$ & -\\
$\eta^0$ & - & $\eta^0\rightarrow \gamma\, \e^+\e^-$ &
$\eta^0 \rightarrow \pi^+\pi^- \e^+\e^-\ (^*)$\\
$\eta'$ & - & $\eta^0\rightarrow \gamma\, \e^+\e^-$ &
$\eta' \rightarrow \pi^+\pi^- \e^+\e^-\ (^*)$\\
$\rho^0$ & $\rho^0\rightarrow \e^+\e^-$ & - & - \\
$\omega^0$ & $\omega^0\rightarrow \e^+\e^-$ &  $\omega^0\rightarrow
\pi^0\, \e^+\e^-$  & -
\\
$\phi^0$ & $\phi^0\rightarrow \e^+\e^-$ & $\phi^0\rightarrow \eta\,
\e^+\e^-\ (^*)$ &
- \\
$J/\psi$ & $J/\psi\rightarrow \e^+\e^-$ &
$J/\psi\rightarrow \gamma\, \e^+\e^-$ & -\\
$\psi'$ & $\psi'\rightarrow \e^+\e^-$ &
$\psi'\rightarrow \gamma \, \e^+\e^-$ & - \\
\D& - & - & $D^\pm \rightarrow \e^\pm \nu_e + X$\\
$B$& - & - & $B^\pm \rightarrow \e^\pm \nu_e + X$\\
\hline
\end{tabular}
\caption{Meson decay channels for di-electron production.}
\label{dilepton_channels_list}
\end{table}

\subsection{Results for central $Pb+Pb$ collisions at \sqrtsnn=2.76~TeV}

\begin{figure}
\includegraphics[width=0.5\textwidth]{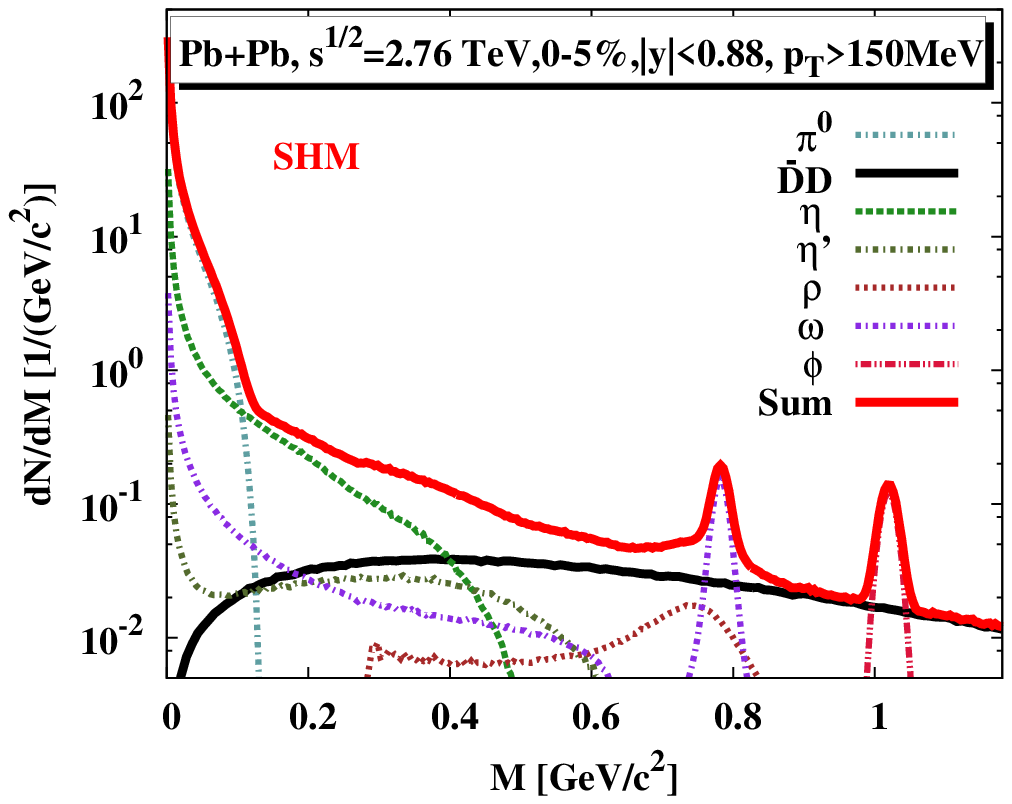}
\includegraphics[width=0.5\textwidth]{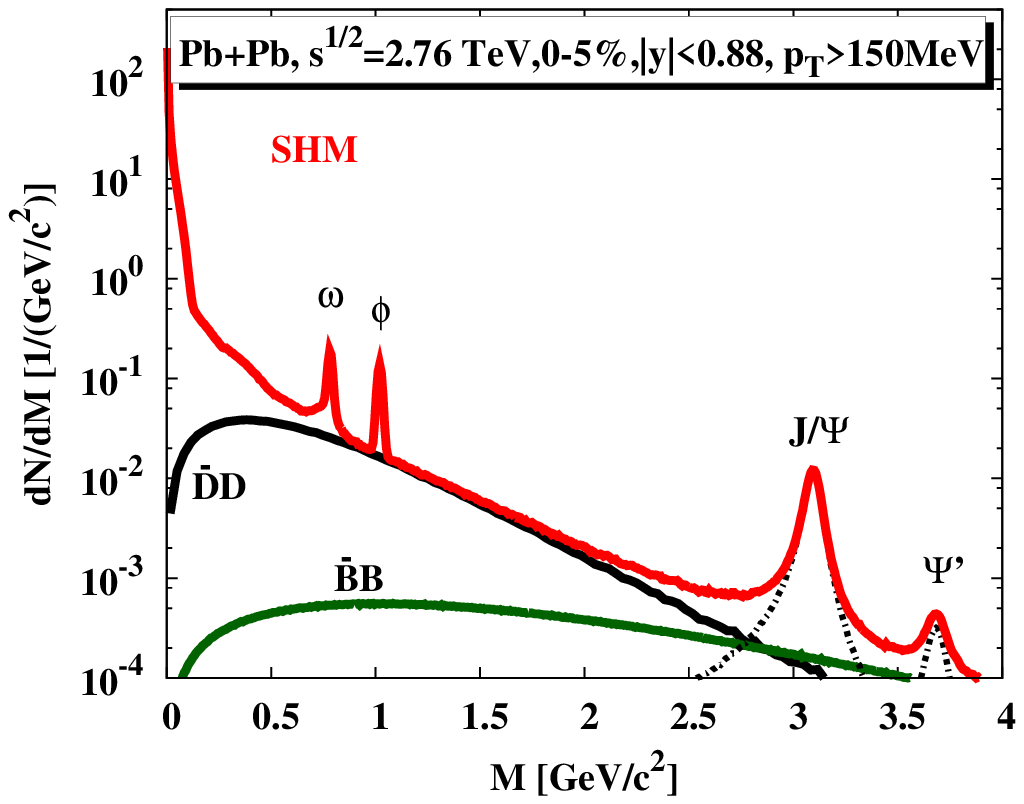}
\caption{(Color on-line) Dielectron
 invariant mass spectrum from various sources evaluated for the 0-5\% most
central $Pb+Pb$ collisions at $\sqrt{s_{NN}}$ = 2.76 TeV within the
SHM {for} the {dielectron invariant} mass range up to {1.2 GeV
(upper window) and} 4 GeV {(lower window)}. {Only} the un-correlated
$D$ and $B$ meson contributions are shown {since almost all
correlations are destroyed in central $Pb+Pb$ collisions}.}
\label{SHM_AA}
\end{figure}

We first present the results for dilepton production from the SHM
at \sqrtsnn=2.76 TeV and then compare to the spectra from the PHSD transport approach,
which incorporates the same cross sections for $c{\bar c}$ as well
as $b {\bar b}$ pair production as specified above, but also
allows for an estimate of hadronic interaction channels as well as
in-medium spectral functions and the radiation from the sQGP. We
concentrate on most central $Pb+Pb$ collisions, since the
uncertainties in the number of participants \NPt~are low and
hadronic in-medium effects are most pronounced.

In Fig.~\ref{SHM_AA} we display our results from the SHM for the
dilepton low mass sector {(upper window)} as well as for the
invariant mass range {up}to 4 GeV {(lower window)}. We recall that
the SHM does not include any dynamical effects and only accounts for
the final hadronic 'cocktail'. The results for di-electron decays of
light mesons are obtained by a scaling of $p+p$ results with the
number of participants as measured by the ALICE Collaboration
\cite{ALICE1}. The only difference between the statistical
hadronization model calculations for the elementary and the
heavy-ion reactions is that $\gamma_S$=0.6 was used for $p+p$
collisions, while $\gamma_S$=0.95 was used in $Pb+Pb$ collisions,
because the latter value better fits the transverse momentum
spectrum of light mesons and multistrange hyperons produced in
$Pb+Pb$ collisions at \sqrtsnn=2.76~TeV.

We point out explicitly that rescattering is fully incorporated in
our calculations for the contributions from open charm and beauty
decays as well as in the dilepton decays of charmonia (in
Fig.~\ref{SHM_AA} and in all subsequent figures in this paper).
{While} heavy-flavor mesons {are} produced fully back-to-back in
$p+p$ collisions (cf. Fig.~\ref{Jaakko276}), we assume that the{se}
initial correlations are washed out {by final-state interactions in
central $Pb+Pb$ collisions since} the survival probability of the
$D$ and $\bar D$ correlation is approximately given by
$R_{AA}(D)^2${, where $R_{AA}$ is the $D$-meson nuclear modification
factor}. {Also, the} energy loss of the produced $D$ and $B$ mesons
was taken into account, both in the correlated and uncorrelated
contributions, according to the Nantes model as described in
Section~\ref{SectNantes}. Additionally, the nuclear modification of
charmonia was taken into account by using the predictions from
Section~\ref{SectTAMU}.

\begin{figure}
\includegraphics[width=0.5\textwidth]{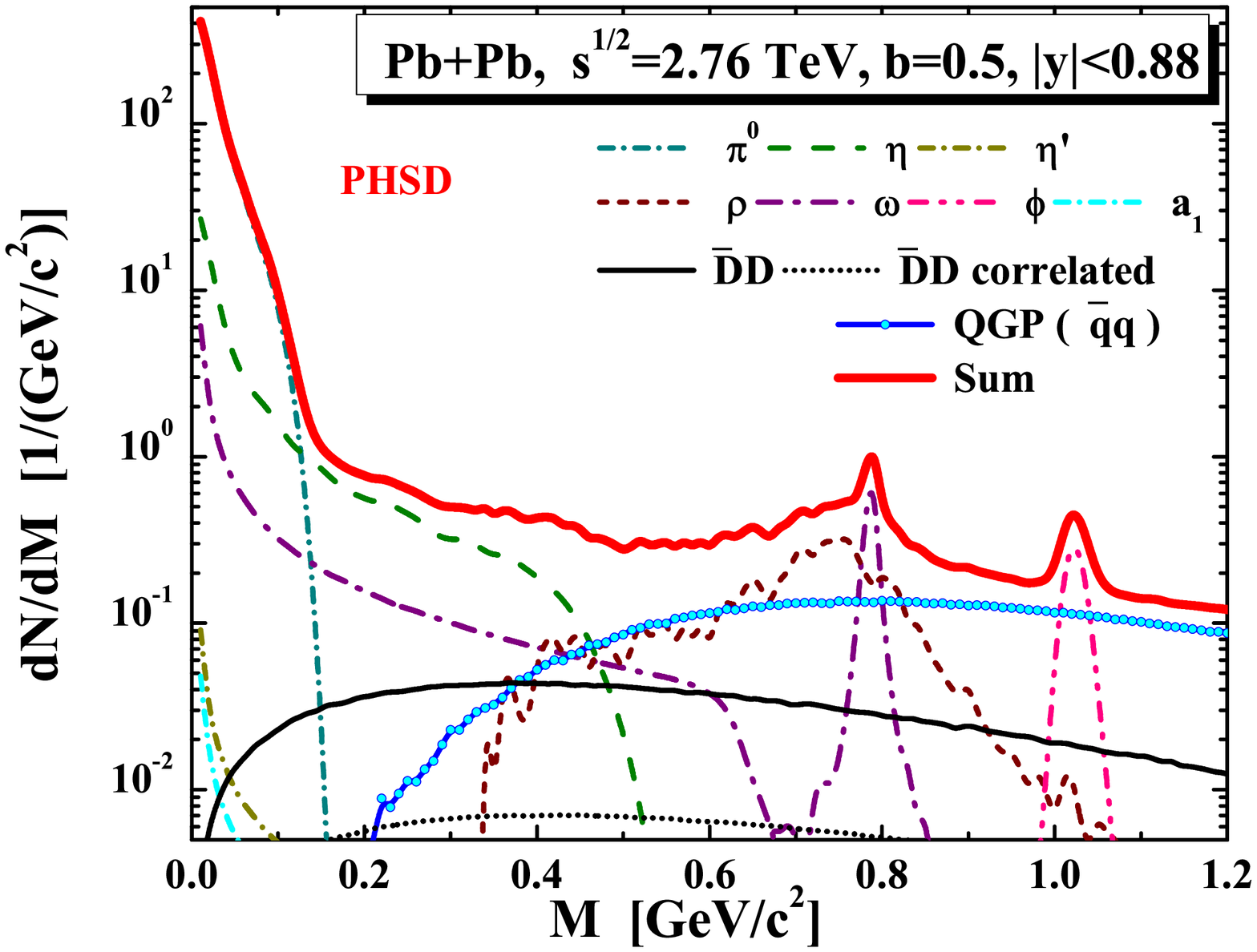}
\includegraphics[width=0.5\textwidth]{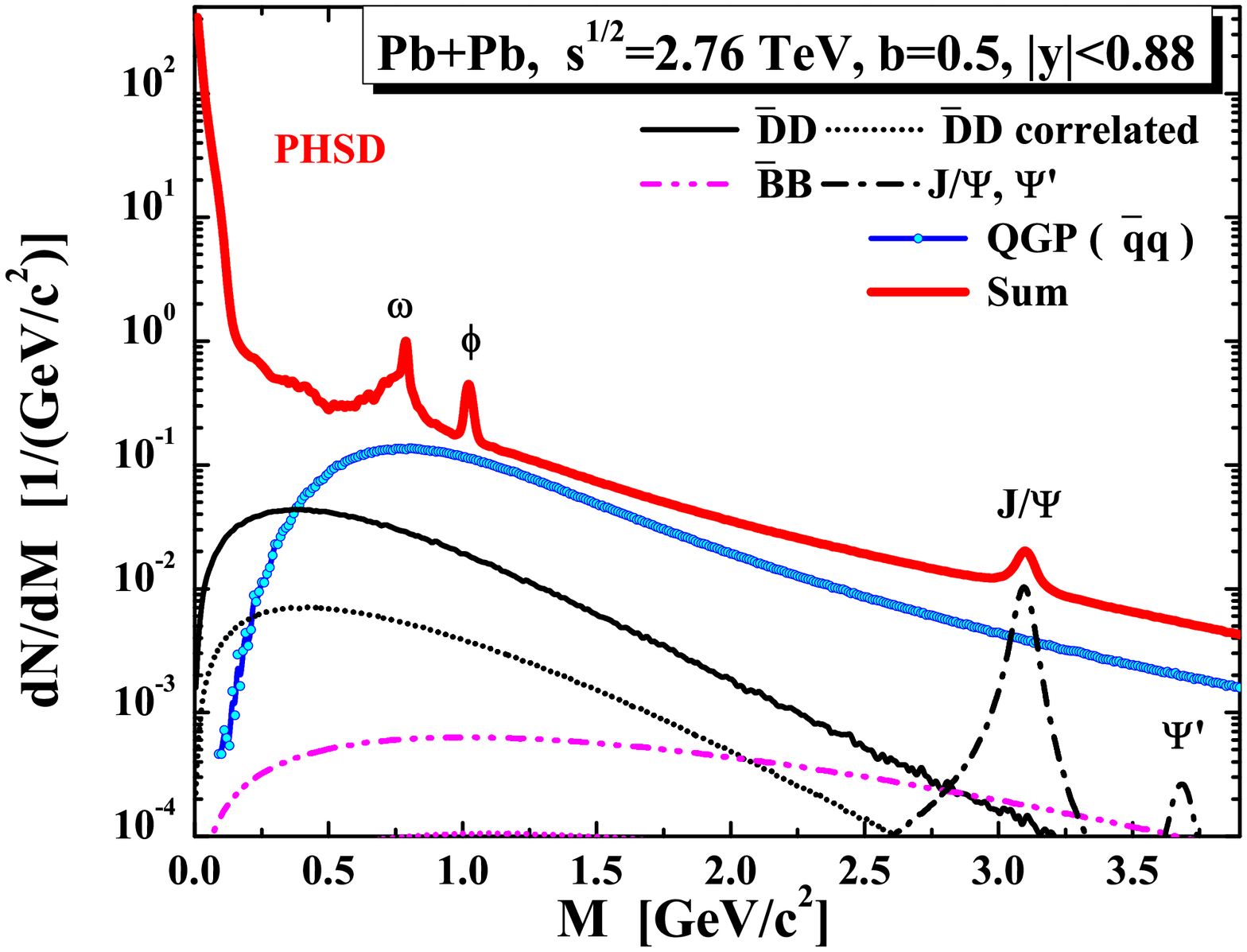}
\caption{(Color on-line) {Same as Fig.\ref{SHM_AA} from the PSHD
with vacuum spectral functions for the vector mesons.} }
\label{PHSD_AA}
\end{figure}

{Results from the} static calculations within the SHM can be
confronted with {those} from the dynamical transport approach PHSD
that are displayed in Fig.~\ref{PHSD_AA}. In general, the low mass
spectrum {from the PHSD} shows a similar channel decomposition as in
the SHM, while the intermediate mass {region} is enhanced
substantially by the partonic radiation channels which are dominated
by the quark-antiquark annihilation channel.  {We note that
although} the contribution of the dileptons from the correlated
decays of $D+\bar D$ and $B+\bar B$ mesons {also peaks at the
intermediate mass region, its}
 total yield is low. {This is due to the fact that
the (partonic) rescattering of the charm and bottom quarks in
central $Pb+Pb$ collisions practically destroys all the correlation
of the initially produced heavy-quark pairs in $p+p$ collisions.} It
is encouraging to see that (Fig.~\ref{PHSD_AA}) the sQGP radiation
is clearly visible in the intermediate mass region{, and this
provides the possibility to measure the dilepton radiation from the
sQGP, making the study of its} properties {experimentally}
accessible. {A similar conclusion was also made in
Ref.~\cite{Linnyk:2011hz} for heavy ion collisions at the top RHIC
energy.}

\begin{figure}
\includegraphics[width=0.5\textwidth]{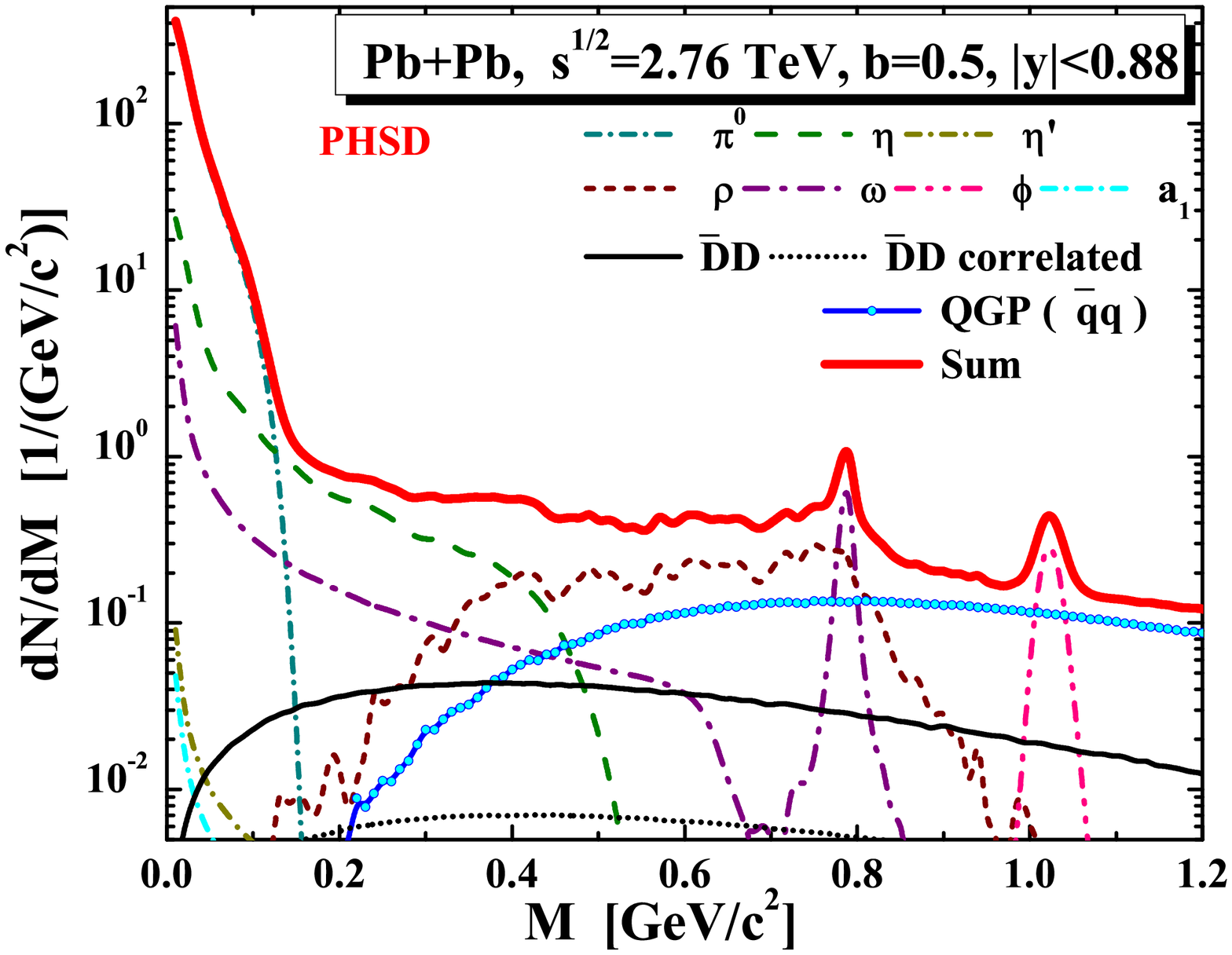}
\includegraphics[width=0.5\textwidth]{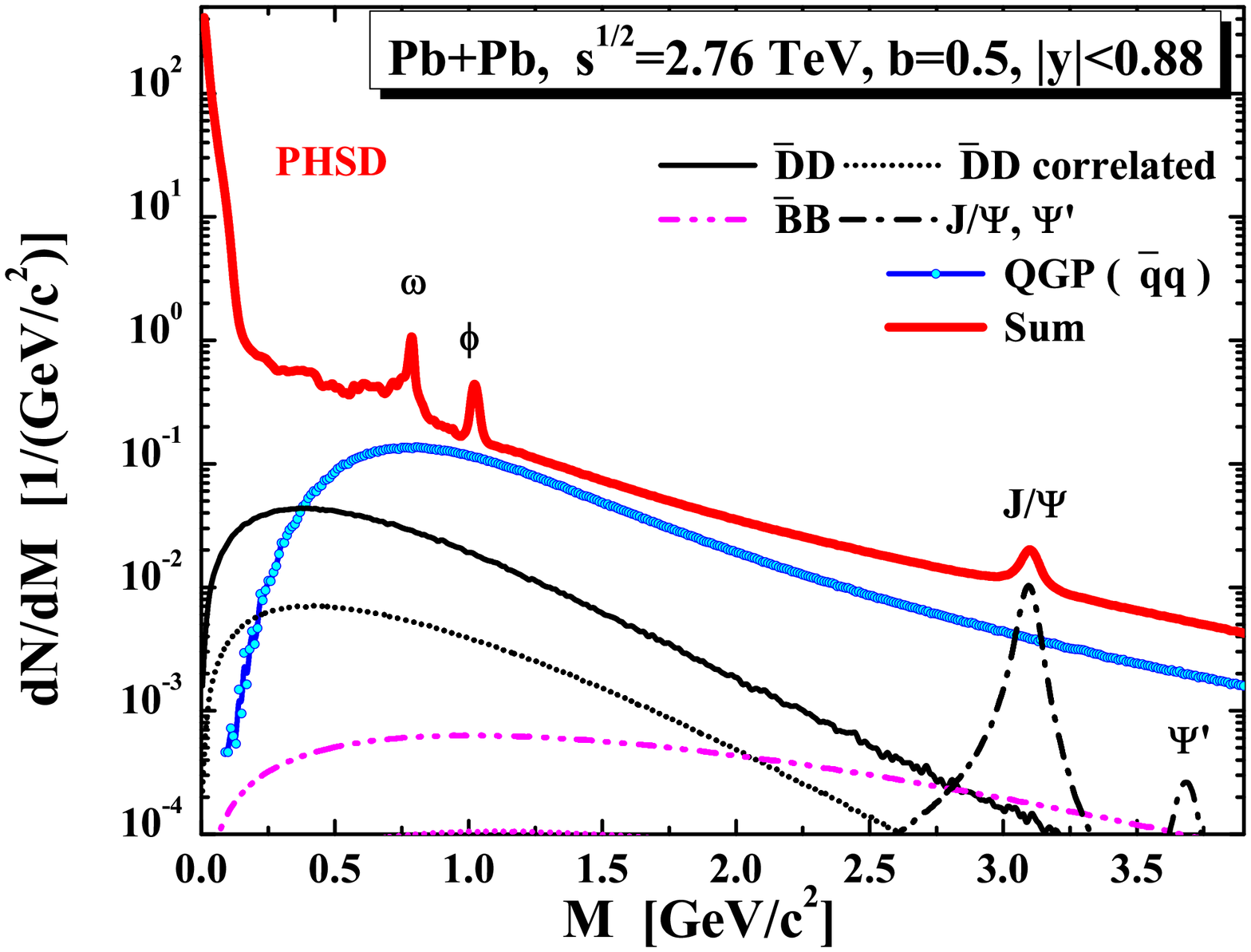}
\caption{(Color on-line) Upper part: Dielectron low invariant mass
spectrum from various different sources evaluated for 10\% central
Pb+Pb collisions at \sqrtsnn=2.76 TeV within the PHSD approach including
in-medium spectral functions for the vector mesons. The lower part
displays our results in the mass range up to 4 GeV.}
\label{PHSD_inmedium}
\end{figure}

\begin{figure}
\includegraphics[width=0.5\textwidth]{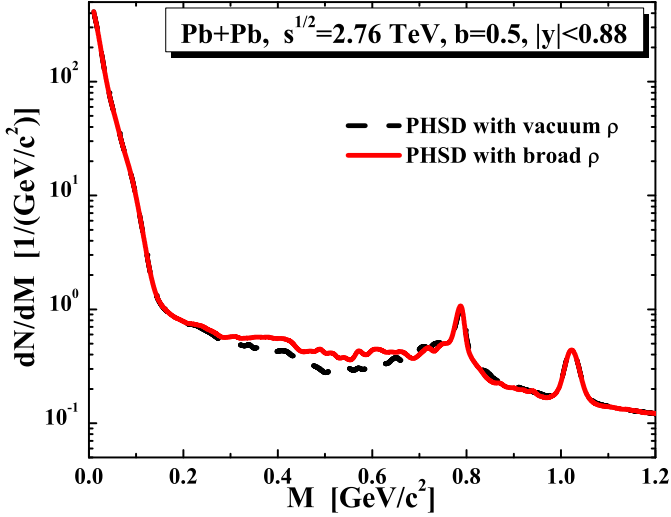}
\caption{(Color on-line) Dielectron low invariant mass spectrum from
various different sources evaluated  for most central Pb+Pb
collisions at \sqrtsnn=2.76 TeV within the PHSD approach with vacuum
as well as with in-medium spectral functions for the vector mesons.
} \label{Compare}
\end{figure}

Another {interesting} question is the modification of the
vector-meson spectral functions by hadronic in-medium effects which
{was} found {to be } essential {for} {describing} the NA60 dilepton
data {in} the low mass regime {from heavy ion collisions at 158
A$\cdot$GeV available from the SPS \cite{Linnyk:2011vx}. In
Fig.~\ref{PHSD_inmedium} we present the dilepton invariant mass
spectrum from the PHSD in the collisional broadening scenario for}
{the} most central Pb+Pb collisions {at \sqrtsnn = 2.76 TeV. It is
seen that the spectra in the intermediate mass region} are not much
{affected by the medium modified vector meson spectral functions}
due to {the} dominance of partonic contributions and correlated
dileptons from $D$-meson decays. A direct comparison of the two
above calculations is displayed in Fig.~\ref{Compare}, where we
focus on the low mass sector. A moderate in-medium effect (within
20\% of the total dilepton yield) in the mass window from 0.3 to
0.6~GeV due to the modification of the vector mesons can be
identified.

\begin{figure}
\includegraphics[width=0.5\textwidth]{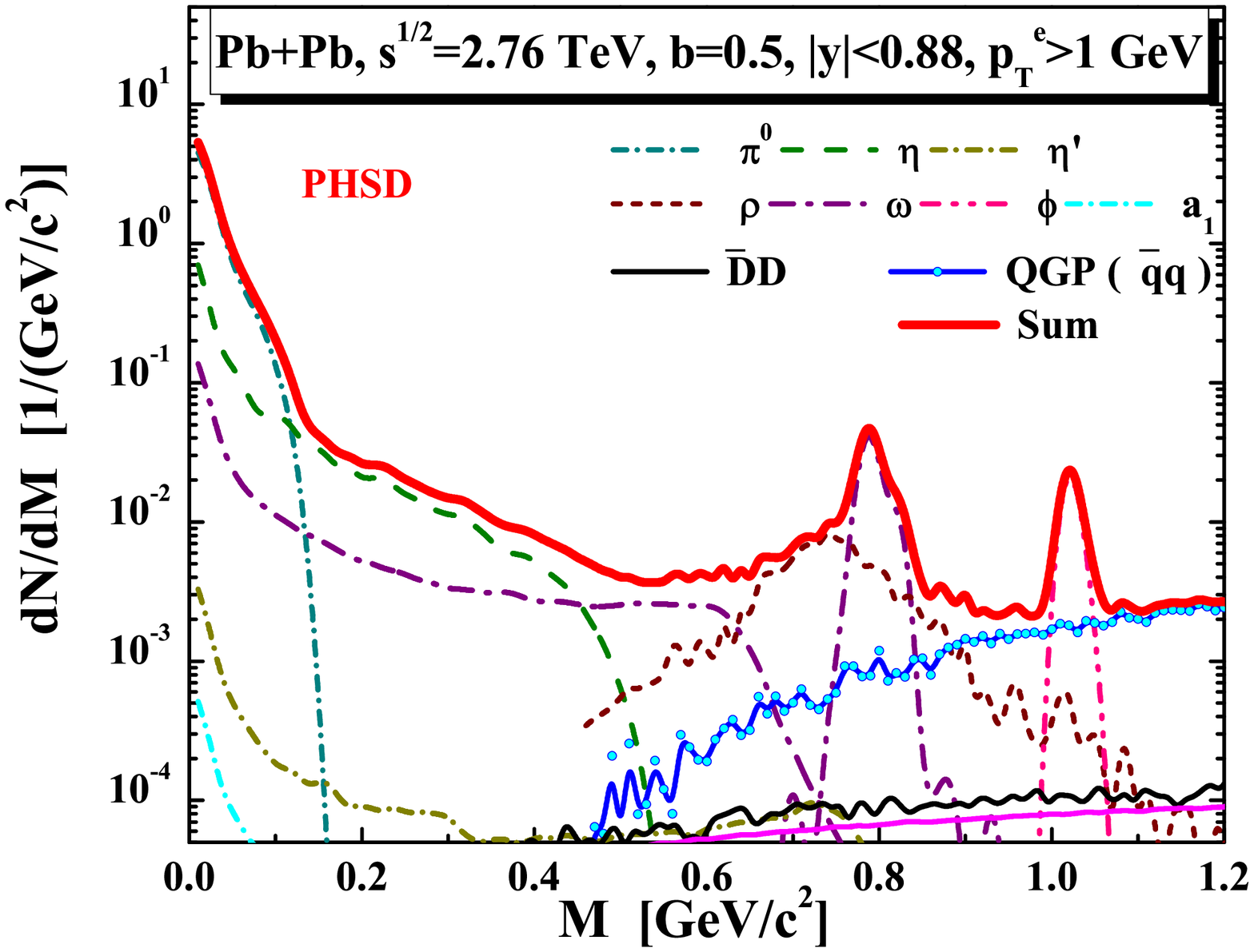}
\includegraphics[width=0.5\textwidth]{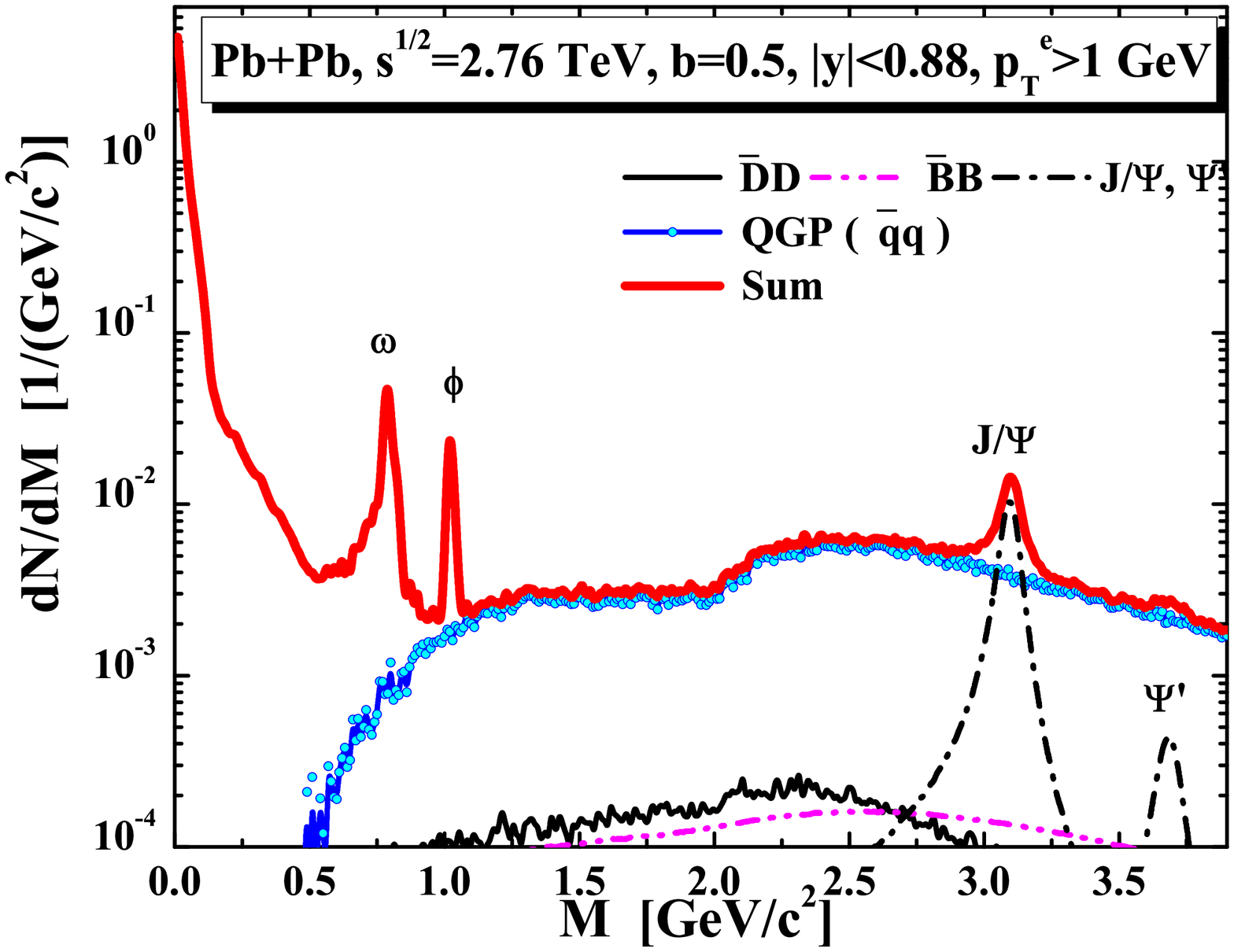}
\caption{(Color on-line) Upper part: Dielectron low invariant mass
spectrum from various different sources evaluated  for the central
Pb+Pb collisions  at \sqrtsnn=2.76 TeV within the PHSD approach including
vacuum spectral functions for the vector mesons with the transverse
momentum cut $p_T > 1$ GeV on the single electron. The lower part
displays our results in the mass range up to 4 GeV.}
\label{PHSD_cut}
\end{figure}

{We next} study the effect of a transverse momentum cut on the
channel decomposition of the dilepton spectrum. {Shown} in
Fig.~\ref{PHSD_cut} {are} the PHSD {results for} the dilepton yield
$dN/dM$ in very central $Pb+Pb$ collisions at \sqrtsnn=2.76 TeV and
mid-rapidity (as in Fig.~\ref{PHSD_AA}) but with the additional cut
on the transverse momentum of single electrons of $p_T>1$ GeV. We
find that the sQGP radiation signal in this case is even more
pronounced {due to the suppression of the} heavy meson contribution
to the dilepton yield {in the} intermediate masses $M=1-3$ GeV. Thus
the $p_T$ cut is beneficial to the extraction of the sQGP signal in
the intermediate mass region.

On the other hand, we see in the upper part of Fig.~\ref{PHSD_cut}
that the low-mass part of the spectrum is strongly modified by the
cut: the $\omega$ and $\phi$ peaks become more prominent while the
total yield is strongly suppressed. This loss of statistics might
impair the detailed measurements of the vector meson spectral shape
and of the relative meson yields. Thus we suggest that, although the
cut on $p_T$ is profitable for the measurement of the intermediate
mass dileptons in order to study the sQGP radiation, the low-mass
spectrum of the dileptons ($M<1.1 GeV$) should better be measured
down to lowest possible values of $p_T$.

\section{Summary}\label{sectionsummary}

In this study{,} we have {extended our previous investigations of
dilepton production at top SPS and top RHIC
energies~\cite{Manninen:2010yf,Linnyk:2011hz,Linnyk:2011vx} to}
 $p+p$ as well as central $Pb+Pb$
collisions at \sqrtsnn=2.76~TeV
{available at the LHC by employing an extended statistical
hadronization model (SHM) and the dynamical PHSD transport
approach.} The SHM serves to independently determine the background
from the 'hadronic cocktail', while the PHSD additionally provides
information on dynamical contributions to the dilepton spectrum,
i.e. a low-mass enhancement due to vector meson modification by
hadronic scatterings as well the contribution from partonic
reactions. Both of the above dynamical effects are not visible in
the final hadronic decay contributions.

In exten{ding} our previous investigations{,} a few new developments
{have been} incorporated apart from a readjustment of the two SHM
parameters ($T$ and $V$) to LHC energies: i) an inclusion of bottom
quarks and antiquarks as well as the corresponding bottom mesons and
their leptonic decay modes, ii) an appropriate model for the energy
loss of open charm mesons in central $Pb+Pb$ collisions
\cite{Gossiaux:2008jv,Gossiaux:2009hr,Gossiaux:2010yx}, and iii) a
dynamical approach for the charmonium suppression and recreation in
the partonic and hadronic phase{s} \cite{Song:2010er,Song:2011xi}
that supersedes the earlier studies within the HSD approach at lower
energies \cite{Olena2}. With respect to the partonic channels for
dilepton production in the PHSD no extensions had to be incorporated
as well as with respect to the overall reaction dynamics of $Pb+Pb$
collisions in the transport approach.

We find a reasonable agreement {between} the differential mass
spectra {of} dileptons from hadronic decays {calculated in} the PHSD
{approach} and the SHM model. Dynamical contributions or medium
effects due to a broadened  $\rho$ spectral function {is seen to}
give a relatively small correction in the low-mass dilepton sector
from $0.3-0.6$~GeV, since the hadronic final state interactions are
less important at LHC energies. However, pronounced traces of the
partonic degrees of freedom are found in the PHSD results {for} the
intermediate mass regime, i.e. between the $\phi$ and $J/\Psi$ decay
peaks{, with the} dilepton {yield} from the strongly interacting
quark-gluon plasma (sQGP) even exceed{ing} that from the
semi-leptonic decays of open charm and bottom mesons. Additionally,
we observe that a transverse momentum cut of 1~GeV/c for the leptons
further suppresses the relative contribution of the heavy meson
decays to the dilepton yield in the intermediate mass region, such
that the sQGP radiation strongly dominates the spectrum for masses
from 1 to 3~GeV, allowing for a closer look at the electromagnetic
emissivity of the partonic plasma in the early phase of Pb+Pb
collisions.

\section*{Acknowledgements}
The authors are grateful for the fruitful discussions with
A.~Andronic, J.~Harris, T.~Hemmick, B.~Jacak, L.~Ruan, J.~Schukraft,
I.~Tserruya, A.~Toia and N.~Xu.
J.M. and E.B. acknowledge financial support through the ``HIC for
FAIR" framework of the ``LOEWE" program.
O.L. acknowledges financial support through the Margaret-Bieber
program of the Justus-Liebig-University of Giessen. The work of
C.M.K. was supported by the U.S. National Science Foundation under
Grants No. PHY-0758115 and No. PHY-1068572, the US Department of
Energy under Contract No. DE-FG02-10ER41682, and the Welch
Foundation under Grant No. A-1358.




\appendix

\section{\D~meson form factors}
\label{apex}

The form factors for $D^0 \rightarrow K^- \e^+ \nu_e$ decays have
been measured by CLEO~\cite{Besson:2009uv},
FOCUS~\cite{Link:2004dh}, BELLE~\cite{Widhalm:2006wz} and
BaBar~\cite{Aubert:2007wg} Collaborations. These can be conveniently
parametrized in the modified pole ansatz~\cite{Becirevic:1999kt}
\begin{equation}
f_+(q^2) =
\frac{f_+(0)}{(1-\frac{q^2}{m_{D_s^*}^2})(1-\alpha\frac{q^2}{m_{D_s^*}^2})}.
\label{f+}
\end{equation}
We take a weighted average $\alpha$=0.35 of the
measured~\cite{Besson:2009uv,Link:2004dh,Widhalm:2006wz,Aubert:2007wg}
$\alpha$ and $m_{D_s^*}$=2.112 GeV as our input for the calculations
and use the same parameters for the corresponding $D^\pm$ decays.

The measured~\cite{Besson:2009uv,Link:2004dh,Widhalm:2006wz} form
factors for $D \rightarrow \pi l \nu_e$ decays can be parametrized
in the simple pole approximation
\begin{equation}
f_+(q^2) = \frac{f_+(0)}{(1-\frac{q^2}{m_{pole}^2})}{,}
\end{equation}
in which we take a weighted average $m_{pole}$=1.92 GeV as input for
our calculations.

The other $D\rightarrow P \e \nu_e$ form factors have not been
measured yet and thus we will rely on theoretical modeling here. We
use the formalism~\cite{Fajfer:2004mv} introduced by Fajfer and
Kamenik in which the form factor for the decay of a \D~meson ($H$)
has the same form as Eq. (\ref{f+}) but the parameters of the
distribution are fixed to the masses of the known (nearest)
resonances. Explicitly, {it is given by}
\begin{equation}
f_+(q^2) =
\frac{f_+(0)}{(1-\frac{q^2}{m_{H}^2})(1-\alpha\frac{q^2}{m_{H}^2})},
\end{equation}
{where} $\alpha=m_{H^*}^2/m_{H^{'*}}^2$. The nearest pole masses
$m_{H^*}$ and $m_{H^{'*}}$ that fix the value of the $\alpha$
parameter are process dependent and are listed in the Table I
of~\cite{Fajfer:2004mv}. The predictions~\cite{Fajfer:2004mv} of the
Fajfer-Kamenik model for the semi-leptonic branching fractions $\D^+
\rightarrow \{\eta', \eta\}\e^+\nu_e$ were recently confirmed by the
measurements~\cite{Mitchell:2008kb,Yelton:2010js} of the CLEO
Collaboration. With the help of the form factors, the $q^2$
distribution of the $D\rightarrow$ pseudo-scalar + $\e +\nu_e$ can
be calculated from
\begin{equation}
\frac{\d \Gamma}{\d q^2} \sim |\vec{p_{X}}(q^2)|^3|f_+(q^2)|^2,
\end{equation}
{where} $X=\{K,\pi,\eta,\eta', f_0\}$. The parameters we have used
for the $D\rightarrow P \e \nu_e$ transitions are listed in Table
\ref{DtoPtran}.

\begin{table}
\begin{tabular}{|c|c|c|c|}
\hline
channel & $m_{pole}$ [GeV] & $\alpha$ & references \\
\hline
$D \rightarrow \pi \e \nu_e$ & 1.92 & 0 & \cite{Besson:2009uv,Link:2004dh,Widhalm:2006wz} \\
$D \rightarrow K \e \nu_e$ & 2.112 & 0.385 & \cite{Besson:2009uv,Link:2004dh,Widhalm:2006wz,Aubert:2007wg} \\
$D \rightarrow \eta \, \e \nu_e$ & 2.010 & 0.74 &  \cite{Fajfer:2004mv} \\
$D \rightarrow \eta' \e \nu_e$ & 2.010 & 0.74 &  \cite{Fajfer:2004mv} \\
$D_s \rightarrow \eta \e \nu_e$ & 2.112 & 0.75 &  \cite{Fajfer:2004mv}\\
$D_s \rightarrow \eta' \e \nu_e$ & 2.112 & 0.75 &  \cite{Fajfer:2004mv}\\
$D_s \rightarrow K \e \nu_e$ & 2.010 & 0.74 & \cite{Fajfer:2004mv}\\
$D_s \rightarrow f_0(980) \e \nu_e$ & 1.7 & 0 & \cite{Ecklund:2009aa}\\
\hline
\end{tabular}
\caption{List of parameters characterizing the $D\rightarrow P \e
\nu_e$ transitions.} \label{DtoPtran}
\end{table}

The $D_{(s)} \rightarrow V \e \nu$  {with} $V$ denot{ing} a vector
meson ($K^*$ or $\phi$) decays can be parametrized with the help of
three form factors~\cite{Bajc:1995km,Fajfer:2005ug}
\begin{eqnarray}
V(q^2)&=&\frac{V(0)}{(1-x)(1-ax)},\nonumber\\
A_1(q^2)&=&\frac{A_1(0)}{1-b'x},\quad A_2(q^2)=\frac{A_2(0)}{1-b'x}
\end{eqnarray}
in the helicity amplitude formalism. {In the a}bove,
$x=q^2/m_{pole}^2$  {with} $m_{pole}$ {being} the (process
dependent) pole meson mass and the free parameters of the model
($V(0)$, $A_1(0)$, $A_2(0)$, $a$ and $b$) were determined
in~\cite{Fajfer:2005ug} (see Table I of~\cite{Fajfer:2005ug} for the
list of $m_{pole}$ and Table II for the values of the fit
parameters){,} and we use these parameters in our calculations.

The $q$-distribution of the total decay rate is evaluated with
\begin{eqnarray}
\frac{\d \Gamma}{\d y} &=& \frac{\d (\Gamma_++\Gamma_-+\Gamma_0)}{\d y} \nonumber \\
&\sim& y|\vec{p_V}(y)|(|H_+(y)|^2+|H_-(y)|^2+|H_0(y)|^2)\nonumber\\
\end{eqnarray}
{where} $y=q^2/m_H^2$ and
\begin{equation}
|\vec{p_V}(y)|^2 = \frac{(m_H^2(1-y)+m_V^2)^2}{4m_H^2}-m_V^2
\end{equation}
is the three momentum of the vector meson in the rest frame of the
\D~meson{,} and the helicity amplitudes read
\begin{eqnarray}
H_\pm(y) &=& (m_H+m_V)A_1(m_H^2y)\mp\frac{2m_H|\vec{p_V}(y)|}{m_H+m_V}V(m_H^2y){,} \nonumber \\
H_0(y) &=& \frac{m_H+m_V}{2m_H m_V \sqrt{y}}(m_H^2(1-y)-m_V^2)A_1(m_H^2 y){,}\nonumber \\
& & - \frac{2m_H|\vec{p_V}(y)|}{m_V(m_H+m_V)\sqrt{y}}A_2(m_H^2 y).
\end{eqnarray}

We have implemented in detail the form factors for each and every of
the known open charm semi-leptonic decay channels. The only decay
channel that is considered isotropic in the momentum space is the $D
\rightarrow \rho^0 \e \nu_e$. Due to the large width of the
$\rho^0$, above formalism is not applicable for this channel. One
could implement the ISGW2 parametrization, but this channel
contributes 3\% of the total semi-leptonic decay width for $D^0$ and
even less for the $D^+${.} {O}mission of {this} form factor should
not {cause much effect}.

\begin{table}
\begin{tabular}{|c|c|c|c|c|c|}
\hline
Hadron & system & Ref. & B [GeV] & n & $p_T$ [GeV] \\
\hline
\jpsi & p+p 2.76 TeV & \cite{Abelev:2012kr} & 4.53 & 4.61 & [0:8] \\
$D$   & p+p 2.76 TeV & \cite{Gossiaux:2008jv,Gossiaux:2009hr,Gossiaux:2010yx} & 2.37 & 3.09 & [0:10] \\
$B$   & p+p 2.76 TeV & \cite{Gossiaux:2008jv,Gossiaux:2009hr,Gossiaux:2010yx} & 6.38 & 3.10 & [0:10] \\
\hline
$D$   & Pb+Pb 2.76 TeV & \cite{Gossiaux:2008jv,Gossiaux:2009hr,Gossiaux:2010yx} & 2.10 & 3.78 & [0:10] \\
$B$   & Pb+Pb 2.76 TeV & \cite{Gossiaux:2008jv,Gossiaux:2009hr,Gossiaux:2010yx} & 4.19 & 3.24 & [0:10] \\
\hline
\end{tabular}
\caption{Parameters characterizing the heavy flavor transverse
momentum spectra around mid-rapidity in $p$+$p$ and $Pb+Pb$
collisions at 2.76 TeV.} \label{Bntable}
\end{table}

\section{Characteristics of the transverse momentum spectra}\label{apexb}

\begin{table}
\begin{tabular}{|c|c|c|c|}
\hline
Hadron      & mass  & relevant branching fractions \\
\hline
$B^+$      & \cite{Amsler:2008zzb} &  $\e^+\nu_e X_c$ 10.2\% \\
$B^0$       & \cite{Amsler:2008zzb} &  $\e^+\nu_e X_c$ 10.1\%  \\
$B^{*+}$   & \cite{Amsler:2008zzb} &  $B^\pm\gamma$ 99\% $B^\pm\e^+\e^-$ 0.5\%~\cite{Landsberg:1986fd} \\
$B^{*0}$     & \cite{Amsler:2008zzb} &  $B^0\gamma$ 99\% $B^0\e^+\e^-$ 0.5\%~\cite{Landsberg:1986fd}\\
\hline
$B_0^+$    &\cite{Godfrey:1986wj,Ebert:1997nk,Bardeen:2003kt,Colangelo:2006kx}   & $B\pi$ \\
$B_0^0$      &\cite{Godfrey:1986wj,Ebert:1997nk,Bardeen:2003kt,Colangelo:2006kx}    & $B\pi$ \\
$B_1^+$    & \cite{Amsler:2008zzb} &  $B^*\pi$ \cite{Isgur:1991wq} \\
$B_1^0$      & \cite{Amsler:2008zzb} &  $B^*\pi$ \cite{Isgur:1991wq}\\
$B_1^{*+}$  &\cite{Godfrey:1986wj,Ebert:1997nk,Bardeen:2003kt,Colangelo:2006kx}  &   $B^*\pi$  \cite{Isgur:1991wq} \\
$B_1^{*0}$   &\cite{Godfrey:1986wj,Ebert:1997nk,Bardeen:2003kt,Colangelo:2006kx}  &   $B^*\pi$ \cite{Isgur:1991wq} \\
$B_2^{*+}$  & \cite{Amsler:2008zzb} &   $B\pi$ 40\% $B^*\pi$ 60\%~\cite{Isgur:1991wq} \\
$B_2^{*0}$    & \cite{Amsler:2008zzb} &   $B\pi$ 40\% $B^*\pi$ 60\%~\cite{Isgur:1991wq} \\
\hline
$B_s$          & \cite{Amsler:2008zzb} &   charm \\
$B_s^{*}$      & \cite{Amsler:2008zzb} &   $B_s^0\gamma$\\
$B_{s0}$      & \cite{Godfrey:1986wj,Ebert:1997nk,Bardeen:2003kt,Colangelo:2006kx,Green:2003zza}   & $BK$\cite{Gorelov:2006fg} \\
$B_{s1}$      & \cite{Amsler:2008zzb} & $B^*K$\cite{Gorelov:2006fg}  \\
$B_{s1}^{*}$  &\cite{Godfrey:1986wj,Ebert:1997nk,Bardeen:2003kt,Colangelo:2006kx,Green:2003zza}  & $B^*K$\cite{Gorelov:2006fg}\\
$B_{s2}^{*}$    & \cite{Amsler:2008zzb}& $B\pi$ 60\% $B^*\pi$ 40\%\\
\hline
\end{tabular}
\caption{References to the works from which we have taken the $B$
meson masses and relevant branching ratios. Unless explicitly stated
otherwise, the branching fractions to the listed channels are
100\%}\label{beautytable}
\end{table}

We do not calculate the transverse momentum spectrum of heavy
hadrons within the SHM but fit a power-law function to the available
experimental data (\jpsi) and theory calculations ($D$ and $B$
mesons) and use these parameterizations as our input for the
simulations. We list here all the relevant parameters related to
transverse momentum spectra of various hadrons.  Explicitly,
\begin{equation}
\frac{\d^2 N}{\d p_T \d y} = C(y)
\frac{p_T}{\Big(1+\frac{p_T^2}{B^2}\Big)^n}{,} \label{powerlawapp}
\end{equation}
 {where} $C(y)$ is some rapidity dependent
normalization factor and the parameters $B$ and $n$ are listed {in
Table \ref{Bntable}} along {with} the references to the input data.

We use the same $p_T$ spectrum for $\psi$' as {that} for \jpsi. In
contrast to the dilepton emission from $D$ and $B$ mesons, our
results do not depend strongly on the details of the
\jpsi~transverse momentum spectrum and hence we use the same
parameters for $Pb+Pb$ collisions as in $p+p$ collisions.

\vspace{10pt}

\section{Properties of open beauty mesons}\label{apexc}

Properties of open beauty mesons are given in
Table~\ref{beautytable}.

\end{document}